\documentclass[10pt,aps,prb,twocolumn,amsmath,amssymb,superscriptaddress,nobibnotes]{revtex4-1}
\usepackage{amsmath}
\usepackage{graphicx}
\usepackage{color}
\usepackage{multirow}
\usepackage[caption = false]{subfig}
\usepackage{array}
\usepackage{threeparttable}
\usepackage[version=4]{mhchem}
\usepackage{mathtools}
\usepackage{siunitx}
\usepackage{centernot}
\usepackage{tikz}
\tikzset{font={\fontsize{11pt}{12}\selectfont}}
\usetikzlibrary{shapes,arrows}
\usepackage{lipsum}
\makeatletter

\renewcommand*{\p@subsection}{}

\renewcommand*{\p@subsubsection}{}
\makeatother

\usepackage{hyperref}
\hypersetup{
	linkcolor=blue,          
	citecolor=blue,        
	urlcolor=blue,           
}

\newcommand{\ket}[1]{|#1\rangle}

\newcommand{\inner}[2]{\langle#1|#2\rangle}

\newcommand{\expecth}[3]{\langle#1|#2|#3\rangle}

\begin{document}
\title{Beyond CCSD(T) accuracy at lower scaling with auxiliary field quantum Monte Carlo}

\author{Ankit Mahajan}
\email{ankitmahajan76@gmail.com}
\affiliation{Department of Chemistry, Columbia 
University, New York, NY 10027, USA}

\author{James H. Thorpe}
\affiliation{Department of Chemistry, Southern Methodist University, Dallas, TX 75275, USA}

\author{Jo S. Kurian}
\affiliation{Department of Chemistry, University of Colorado, Boulder, CO 80302, USA}

\author{\\David R. Reichman}
\affiliation{Department of Chemistry, Columbia University, New York, NY 10027, USA}

\author{Devin A. Matthews}
\affiliation{Department of Chemistry, Southern Methodist University, Dallas, TX 75275, USA}

\author{Sandeep Sharma}
\email{sanshar@gmail.com}
\affiliation{Department of Chemistry, University of Colorado, Boulder, CO 80302, USA}

\begin{abstract}
We introduce a black-box auxiliary field quantum Monte Carlo (AFQMC) approach to perform highly accurate electronic structure calculations using configuration interaction singles and doubles (CISD) trial states. This method consistently provides more accurate energy estimates than coupled cluster singles and doubles with perturbative triples (CCSD(T)), often regarded as the gold standard in quantum chemistry. This level of precision is achieved at a lower asymptotic computational cost, scaling as \(O(N^6)\) compared to the \(O(N^7)\) scaling of CCSD(T). We provide numerical evidence supporting these findings through results for challenging main group and transition metal-containing molecules. 
\end{abstract}
\maketitle

\section{Introduction}





Phaseless auxiliary field quantum Monte Carlo (AFQMC), a fermionic projector Monte Carlo (PMC) technique developed by Zhang and coworkers \cite{zhang1995constrained,zhang1997constrained,zhang2003quantum,motta2018ab}, has attracted significant attention in quantum chemistry. Its accuracy and computational cost largely depend on the choice of the trial wave function used to control the sign (or phase) problem inherent to fermionic PMC techniques. When a single Hartree-Fock (HF) determinant is used as the trial state (AFQMC/HF), its cost scales as \(O(N^5)\) for a fixed stochastic error in the total energy. Despite this relatively moderate scaling, AFQMC/HF has been shown to provide more accurate energies than coupled cluster singles and doubles (CCSD),\cite{lee2022twenty} which has a computational scaling of \(O(N^6)\). But the same benchmark study\cite{lee2022twenty} found AFQMC/HF to be less accurate than the coupled cluster singles and doubles with perturbative triples (CCSD(T))\cite{raghavachari1989fifth} method when calculating atomization energies for weakly correlated systems. 

Since AFQMC becomes formally exact in the limit of the exact trial, a straightforward way to improve its accuracy is to use a more accurate trial. Multideterminant states are often employed in AFQMC for this purpose, usually obtained from complete active space (CAS) calculations, where the active orbital spaces are chosen based on chemical intuition.\cite{al2007bond,purwanto2015auxiliary,shee2019achieving,williams2020direct} 
The approximation quality can critically depend on the somewhat subjective choice of the active space, a trait shared by all CAS-based approaches. Recently, efforts have been made to remedy this using automatically chosen active spaces, based on natural orbitals or other criteria.\cite{malone2022ipie,neugebauer2023toward,wei2024scalable} Despite improvements in the ability to automatically select the multideterminant trial states, all these studies are limited by the high cost of the calculations. This is ameliorated to a large extent by the development of efficient algorithms that enable the use of long expansions--up to a million determinants--as trial states in AFQMC.\cite{mahajan2020efficient,mahajan2021taming,mahajan2022selected} The development of these algorithms has provided a means to systematically improve the accuracy of AFQMC energies by using larger active spaces and selected configuration interaction (sCI) expansions.\cite{malone2022ipie,wei2024scalable,jiang2024improved} However, as with other systematically improvable methods including density matrix renormalization group\cite{white1999ab}, various flavors of sCI\cite{Huron1973,Yann2017,Holmes2016b,ShaHolUmr,tubman2016deterministic} and other multireference methods, AFQMC with sCI trial states is not a polynomial scaling black-box algorithm. 

Although these systematically improvable methods can be beneficial for solving specific challenging problems, in the context of computational chemistry, a black-box method with polynomial scaling is often preferable. There have been some attempts to improve upon AFQMC/HF in such a way. These have mainly included the use of symmetry-broken and projected mean field trial states.\cite{shi2014symmetry,lee2020utilizing,xiao2021pseudo,danilov2024capturing} Another technique is the self-consistent optimization of the mean-field trial state using the one-body density matrix.\cite{qin2016coupling} Both these approaches have the benefit of preserving the cost scaling of AFQMC/HF, but have yet to be thoroughly benchmarked for \textit{ab initio} systems. 

In this work, we introduce a new black-box method by using the configuration interaction singles and doubles (CISD) wave function as the trial state. The resulting algorithm is denoted as AFQMC/CISD and it consistently yields highly accurate ground state energies. Leveraging the algorithmic developments for handling long determinant expansions, this method has a computational cost scaling of \(O(N^6)\) for a fixed stochastic error in the total energy. Our benchmark calculations demonstrate that this approach provides energies more accurate than CCSD(T), despite its lower cost scaling. The test cases include both main group and transition metal containing systems with varying levels of correlation. 

We have implemented this method in a code optimized for GPUs, which significantly mitigates the large prefactor associated with the cost scaling of QMC methods.\cite{shee2018phaseless,malone2020accelerating,malone2022ipie,jiang2024improved,huang2024gpu} Due to its lower cost scaling and the use of GPUs, we find that the practical walltime cost of AFQMC/CISD  for obtaining reasonable stochastic errors is comparable to that of a CPU implementation of CCSD(T) using roughly equivalent computational resources (see Section \ref{sec:cost} for details). As an example, for the ground state energy of polyacetylene \ce{C12H14} in a double-\(\zeta\) basis, we found the walltime cost of the UCCSD(T) calculation to be 8 hours using 30 CPU cores compared to the AFQMC/CISD cost of 13 hours on a single GPU for a 1 mH stochastic error. 

Similar to the hierarchies in configuration interaction or coupled cluster methods, the accuracy of AFQMC can be improved by including higher-order excitations beyond doubles in the trial state. Remarkably, the cost of using these trial states in AFQMC is less than or equal to the cost of obtaining them by solving the corresponding CI or CC equations. For example, the cost of solving CCSD/CISD, CCSDT/CISDT, and CCSDTQ/CISDTQ equations scales as \(O(N^6)\), \(O(N^8)\), and \(O(N^{10})\), respectively. But using the truncated CI wave functions of the same order as trial states in AFQMC has a cost of \(O(N^6)\), \(O(N^7)\) and \(O(N^{9})\), respectively (see section~\ref{sec:wicks} for more details). It should, however, be noted that very often the bottleneck in performing these higher-order calculations is the large memory required to store the wave functions, which remains a limiting factor. The cost of storing doubles, triples, and quadruples CC or CI amplitudes as dense tensors scales as \(O(N^4)\), \(O(N^6)\), and \(O(N^8)\), respectively.

This paper is organized as follows: We first present the theory and computational details of AFQMC/CISD in section \ref{sec:theory}. In section \ref{sec:results}, we present benchmark results for main group (\ref{sec:main_group}) and transition metal containing molecules (\ref{sec:transition_metal}). We present an analysis of the scaling of ground state energy and computational cost of the method in comparison to CCSD(T) in section \ref{sec:cost}. We conclude with a summary of the results and future directions in section \ref{sec:conclusion}.

\section{Theory}\label{sec:theory}
\begin{table}[t]
   \begin{tabular}{cl}
   \hline 
   Symbol & Meaning\tabularnewline
   \hline 
   $N$ & number of electrons\tabularnewline
   $M$ & number of basis functions\tabularnewline
   $X$ & number of Cholesky vectors\tabularnewline
   $i,j,\dots$ & general orbital indices\tabularnewline
   $p,q,\dots$ & occupied orbital indices\tabularnewline
   $t,u,\dots$ & virtual orbital indices\tabularnewline
   \hline 
   \end{tabular}
   \caption{Glossary of the notation used in this article.}\label{tab:symbols}
\end{table}

We use the quantum chemistry Hamiltonian given by
\begin{equation}
	H = \sum_{ij} h_{ij}a_i^{\dagger}a_j + \frac{1}{2}\sum_{\gamma}\left(\sum_{ij}L^{\gamma}_{ij}a_i^{\dagger}a_j\right)^2,\label{eq:ham}
\end{equation}
where \(h_{ij}\) are modified one-electron integrals and \(L^{\gamma}_{ij}\) are Cholesky decomposed two-electron integrals in an orthonormal orbital basis, and \(\{a_i^{\dagger}\}\) and \(\{a_i\}\) are electronic creation and annihilation operators, respectively. We denote the number of electrons by \(N\), the number of orbitals by \(M\), and the number of Cholesky vectors by \(X\). Table \ref{tab:symbols} shows a summary of the notation used in this paper. In cost scaling expressions that make a distinction between \(X, M\) and \(N\), we assume \(X > M \gg N\). When such a distinction is not made, \(N\) is used as a proxy for the system size. The ground state is obtained by applying an exponential form of the projector onto an initial state, \(\ket{\phi_0}\), usually taken to be the Hartree-Fock (HF) determinant, as
\begin{equation}
   e^{-\tau \hat{H}}\ket{\phi_0} \xrightarrow{\tau\rightarrow\infty}\ket{\Psi_0},
\end{equation}
where \(\tau\) is the imaginary time, \(\ket{\Psi_0}\) is the ground state, and we assume \(\inner{\phi_0}{\Psi_0}\neq 0\). In AFQMC, the action of the projector (also referred to as propagator) is sampled using auxiliary fields, and the ground state is represented statistically as a weighted sum of non-orthogonal Slater determinants
\begin{equation}
   \ket{\Psi_0} = \sum_{\mu} w_{\mu} \frac{\ket{\phi_{\mu}}}{\inner{\psi_T}{\phi_{\mu}}},
\end{equation}
where \(w_{\mu}\) are weights, \(\ket{\phi_{\mu}}\) are walker Slater determinants, and \(\psi_T\) is the trial state used for importance sampling. Details of this procedure can be found in Ref. \citenum{motta2018ab}. We note that this representation of the ground state is biased due to the phaseless approximation\cite{zhang2003quantum} employed during propagation to control the sign or phase problem. The extent of this bias is determined by the trial state (\(\ket{\psi_T}\)) used in the phaseless approximation.\cite{zhang2003quantum} 
\subsection{CISD trial states}
In this paper, we use CISD trial states given by
\begin{equation}
   \ket{\psi_T} = \left(1 + \sum_{pt}c^{t}_{p}a_{t}^{\dagger}a_{p} + \frac{1}{2}\sum_{ptqu}c^{tu}_{pq}a_{t}^{\dagger}a_{u}^{\dagger}a_{q}a_{p}\right)\ket{\phi_0},        
\end{equation}
where we have used spin orbitals for notational brevity. We use the singles and doubles CI coefficients based on CCSD amplitudes as 
\begin{equation}
   c^{t}_{p} = \tau^{t}_{p},\quad c^{tu}_{pq} = \tau^{tu}_{pq} + \tau_{p}^{t}\tau_{q}^{u},
\end{equation}
where \(\tau\) are the canonical CCSD amplitudes. Other potential sources for the CI coefficients include variational CISD, externally corrected CCSD,\cite{paldus1994valence} and sCI. In our calculations, we found that CCSD and variational CISD-based coefficients yielded roughly similar AFQMC energies, with the former showing greater accuracy in some instances. A detailed comparison of these options will be addressed in future studies. 

We consider the evaluation of three trial and walker dependent quantities: the overlap of the walker with the trial state, \(\inner{\psi_T}{\phi}\), along with the force bias,\cite{zhang2003quantum} defined as
\begin{equation}
   v_{\gamma} = \sum_{ij}L^{\gamma}_{ij}\frac{\expecth{\psi_T}{a_i^{\dagger}a_j}{\phi}}{\inner{\psi_T}{\phi}},
\end{equation}  
and the local energy, given by
\begin{equation}
   E_{L} = \frac{\expecth{\psi_T}{H}{\phi}}{\inner{\psi_T}{\phi}}.
\end{equation}
The force bias is a dynamic generalization of a static integral contour deformation\cite{rom1997shifted} and serves to minimize the fluctuations in the importance sampling function. The weighted average of the local energy provides the mixed estimator of the desired energy. We present two ways to calculate these quantities, one based on automatic differentiation and the other based on direct evaluation of the expressions using Wick's theorem. 
\begin{figure*}[t]
   \centering
   \includegraphics[width=0.8\textwidth]{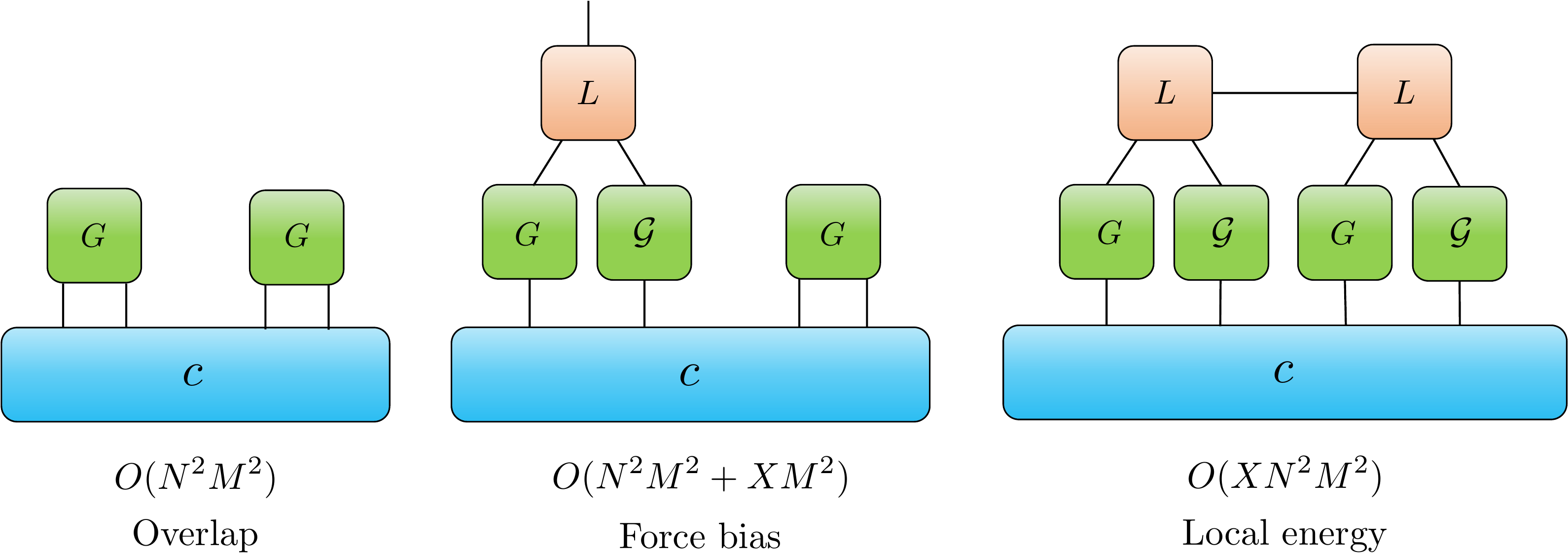}
   \caption{Tensor diagrams for the most expensive terms in the evaluation of overlap, force bias, and local energy for a CISD trial state. Computational cost scaling is shown below each diagram.}
   \label{fig:cisd}
\end{figure*}
\subsubsection{Using overlap derivatives}\label{sec:ad}
The use of derivatives to efficiently evaluate quantities such as the local energy was first introduced by Filippi \textit{et al.} in the context of real space QMC.\cite{filippi2016simple} A related approach for AFQMC in orbital space was reported in Ref. \citenum{jiang2024unbiasing}. Below we outline the key aspects of this technique that can be applied to a general trial state. The method only requires the ability to compute the overlap between the walker and the trial. For a CISD state, the overlap is calculated using the generalized Wick's theorem\cite{balian1969nonunitary} as
\begin{equation}
   \frac{\inner{\psi_T}{\phi}}{\inner{\phi_0}{\phi}} = 1 + \sum_{pt}c^{t}_{p}G_{t}^{p} + \frac{1}{2}\sum_{ptqu}c^{tu}_{pq}(G_{t}^{p}G_{u}^{q} - G_{u}^{p}G_{t}^{q}),
\end{equation}
where \(G\) is the Green's function, given by
\begin{equation}
   G_{j}^{i} = \frac{\expecth{\phi_0}{a_i^{\dagger}a_j}{\phi}}{\inner{\phi_0}{\phi}} = \left(U(V^{\dagger}U)^{-1}V^{\dagger}\right)^{j}_{i},
\end{equation}
with 
\begin{equation}
   \inner{\phi_0}{\phi} = \det(U^{\dagger}V).
\end{equation}
Here \(U\) and \(V\) are the orbital coefficient matrices of the walker and the CISD reference HF state, respectively. Note that the cost of overlap evaluation  ($\inner{\psi_T}{\phi}$) scales as \(O(N^2M^2)\). The force bias can be written as 
\begin{equation}\label{eq:force_bias_auto}
   v_{\gamma} = \frac{\partial}{\partial x_{\gamma}}\frac{\expecth{\psi_T}{\exp\left(\sum_{\gamma' ij}x_{\gamma'}L^{\gamma'}_{ij}a_i^{\dagger}a_j\right)}{\phi}}{\inner{\psi_T}{\phi}}\bigg\vert_{\mathbf{x}=0}.
\end{equation} 
Note that the operator in the exponent in the numerator is a one-body operator. Therefore, by applying Thouless' theorem, its action on the walker determinant generates another Slater determinant. The force bias can then be calculated as the gradient of the overlap between this rotated walker and the trial. Using reverse mode automatic differentiation, the gradient can be obtained at the same cost scaling as the overlap itself, and the resulting cost of evaluating the force bias for a CISD trial is \(O(N^2M^2 + XM^2)\).

The one-body part of the local energy can be evaluated similarly to the force bias. We write the two-body part as
\begin{equation}
   E_L^2 = \frac{1}{2}\frac{\partial^2}{\partial x^2}\sum_{\gamma}\frac{\expecth{\psi_T}{\exp\left(x\sum_{ij}L^{\gamma}_{ij}a_i^{\dagger}a_j\right)}{\phi}}{\inner{\psi_T}{\phi}}\bigg\vert_{x=0}.
\end{equation}
Again, due to Thouless' theorem, the numerator reduces to an overlap between the trial and a Slater determinant. Because the derivative involves a single variable, it can be evaluated using automatic differentiation or finite difference at the same cost scaling, given by \(O(XN^2M^2)\) for the CISD trial state.

Even though force bias and overlap calculations are carried out more frequently than local energy evaluation during an AFQMC run--typically about 50 times more often--energy evaluation becomes the bottleneck for larger systems in practice due to its higher asymptotic scaling. Since the variance in the energy grows roughly linearly with the number of electrons, the asymptotic cost to obtain a fixed stochastic error with AFQMC/CISD scales as \(O(XN^3M^2)\), which is effectively of the order of the sixth power of the system size.

\subsubsection{Explicit evaluation using Wick's theorem}\label{sec:wicks}
An alternative to the derivative-based approach exists for multideterminantal trial states, such as CISD. This technique makes use of the particle-hole excitation structure of the trial state to efficiently compute various quantities. The details of this approach are outlined in Refs. \citenum{mahajan2021taming} and \citenum{mahajan2022selected}. Briefly, it proceeds by explicitly evaluating the expressions for the force bias and local energy using the generalized Wick's theorem. Specifically, we make use of the identity
\begin{equation}\label{eq:eloc_wick}
   \frac{\expecth{\phi_0}{\left(\prod_{\mu}^k a^{\dagger}_{p_{\mu}}a_{t_{\mu}}\right)a_i^{\dagger}a_k^{\dagger}a_la_j}{\phi}}{\inner{\phi_0}{\phi}} = \det \begin{pmatrix}
		 G^{\left\{i,k\right\}}_{\left\{j,l\right\}} & \mathcal{G}^{\left\{i,k\right\}}_{\left\{t_{\mu}\right\}}\\[1em]
		 G^{\left\{p_{\mu}\right\}}_{\left\{j,l\right\}} & G^{\left\{p_{\mu}\right\}}_{\left\{t_{\mu}\right\}}\\
	 \end{pmatrix},
\end{equation}
where the operator in parentheses on the LHS is a particle-hole excitation operator of rank \(k\), the notation using sets of indices like \(G^{\{p_{\mu}\}}_{\{t_{\mu}\}}\) denote slices of the Green's function matrix, and the modified Green's function is given as
\begin{equation}
   \mathcal{G}^i_j = G^i_j - \delta^i_j.
\end{equation}   
The complete expressions for the force bias and local energy for a CISD trial state are provided in Appendix \ref{app:cisd}. The most computationally expensive terms are shown in Fig. \ref{fig:cisd} as tensor diagrams, alongside their associated cost scaling. Notably, the cost scaling of this approach matches that of the derivative-based approach for all relevant quantities. This no longer holds when higher-order CISDT and CISDTQ trial states are used. The most expensive terms in these cases are shown in Fig. \ref{fig:cisdtq} of Appendix \ref{app:cisd}. By constructing appropriate intermediate quantities, these can be evaluated with cost scalings of \(O(XN^2M^2 + N^3M^3)\) for CISDT and \(O(N^4M^4)\) for CISDTQ. In contrast, the derivative-based approach results in higher cost scalings of \(O(XN^3M^3)\) for CISDT and \(O(XN^4M^4)\) for CISDTQ trial states.

In our implementation of these two algorithms for performing AFQMC/CISD, we found that the second approach using explicit Wick's theorem expressions was up to 5 times faster than the derivative-based approach depending on the system size. 
For example, for a polyacetylene chain containing 10 carbon atoms in the aug-cc-pVDZ basis, the second approach was roughly 5 times faster for AFQMC/CISD calculations based on a UHF (unrestricted HF) reference. Nonetheless, the ease of implementation of the derivative-based method within an AD framework makes it very useful for quick prototyping and debugging. 

\section{Results}\label{sec:results}
In this section, we present benchmark results for the accuracy and computational cost of AFQMC/CISD. The analysis covers main group molecules from the HEAT dataset,\cite{tajti2004heat} a subset of the W4 dataset,\cite{karton2017w4} and transition metal-containing molecules, which are known to be particularly challenging for many electronic structure methods. We make comparisons with various coupled cluster methods including CCSD, CCSD(T), and CCSDTQ. Hartree-Fock, CCSD, and CCSD(T) calculations were performed using PySCF,\cite{sun2018pyscf} while CCSDTQ calculations were done with MRCC\cite{kallay2020mrcc} and CFOUR\cite{matthews2020coupled} programs for open-shell and closed-shell molecules, respectively. We did all AFQMC calculations using our AD-AFQMC code available in a public repository.\cite{dqmc_code} This python code is written using the Jax library,\cite{jax2018github} and supports both CPU and GPU computations. We used a single NVIDIA A100 GPU along with four AMD Milan CPU cores for all of the AFQMC calculations. 

We use UHF-based trial states for both AFQMC/HF and AFQMC/CISD calculations unless explicitly stated otherwise. The trial CISD wave functions are obtained by truncating the unrestricted CC wave function on top of a UHF reference state. We use restricted initial walkers, with orbitals taken to be the \(\alpha-\)spin UHF orbitals, to enforce spin symmetry in AFQMC.\cite{purwanto2008eliminating} This choice was found to minimize systematic errors in general. For CC calculations, in most cases, RHF-based methods were found to be more accurate. Therefore, we have generally used RHF-based CC methods, except for some cases where the alternative choice is explicitly specified. The frozen-core approximation was used in all calculations without pseudopotentials: we froze the He core for second-row atoms, Ne core for third-row atoms, and Ar core for the \(3d\) transition metal atoms (except for \ce{[Cu2O2]^2+}). This allows us to use a time step of 0.005 a.u. in AFQMC propagation.\cite{wei2024scalable} Note that UHF-based CC methods use an unrestricted core, whereas in AFQMC we freeze a restricted core. We expect discrepancies due to this difference in the treatment of frozen orbitals, both in CC energies and when CC amplitudes are used to make the CISD trial states, to be very small because of the tight cores used. A cutoff of $10^{-5}$ was used for calculating the modified Cholesky integrals used in AFQMC. We obtain correlation energies in the continuum limit using a two-point extrapolation based on the inverse cubic relation\cite{helgaker1997basis}
\begin{equation}
   E_{\text{corr}}^X = E_{\text{corr}}^{\infty} + \frac{a}{X^3},
\end{equation}
where \(E_{\text{corr}}^X\) is the correlation energy in a basis set with cardinal number \(X\), and \(E_{\text{corr}}^{\infty}\) is the correlation energy in the basis set limit. We perform stochastic error propagation using the bootstrap sampling approach.\cite{becca2017quantum}

\begin{figure*}[t]
   \centering
   \includegraphics[width=0.9\textwidth]{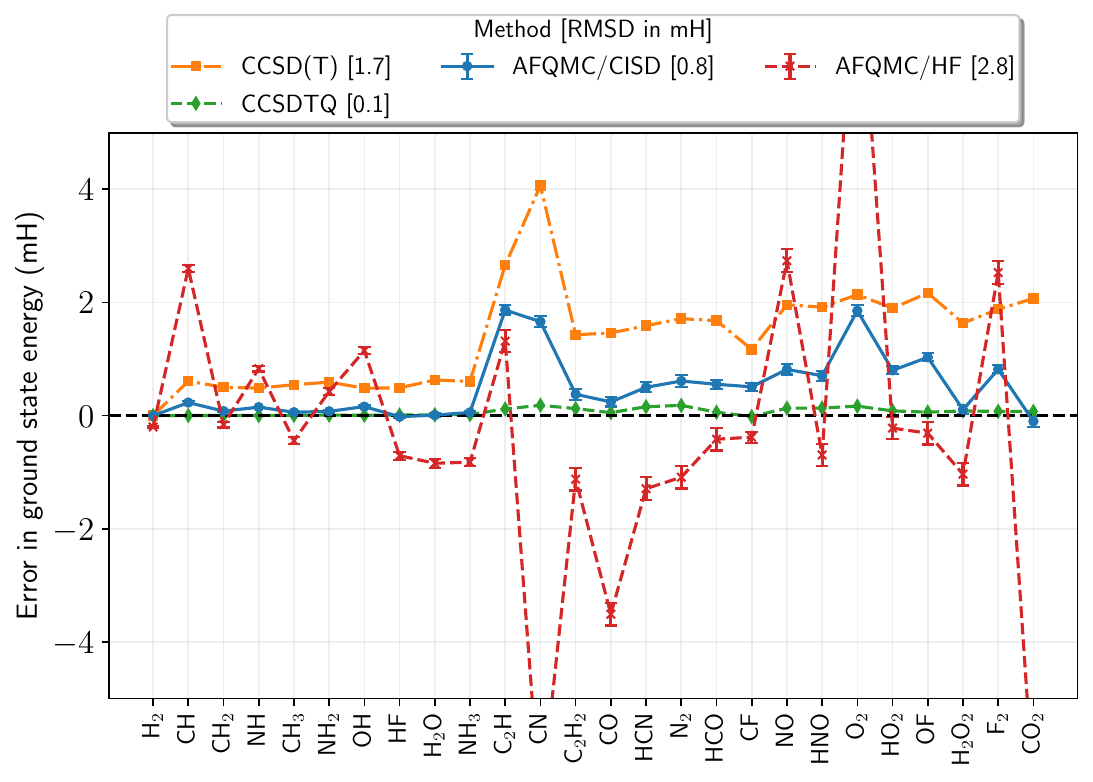}
   \caption{Errors in the ground state energies of molecules in the HEAT dataset (cc-pVDZ basis) calculated using different methods. The errors are with respect to CCSDTQP taken from Ref. \citenum{bomble2005coupled}. AFQMC/HF energies are from Ref. \citenum{sukurma2023benchmark}}
   \label{fig:heat}
\end{figure*}
\subsection{Main group molecules}\label{sec:main_group}
We begin examining the performance of various methods in molecules containing the main group elements. We test the performance for weakly correlated (HEAT dataset and isomerization energies for the W4-ISO20 subset), moderately strongly correlated (W4-MR dataset), and significantly strongly correlated systems (\ce{N2} dissociation). We also test the accuracy of calculated dipole moments for a set of small molecules. 

\subsubsection{HEAT dataset}
The HEAT dataset, developed by Stanton and colleagues, contains small open and closed shell molecules made of light elements.\cite{tajti2004heat} It aims to provide highly accurate theoretical atomization energies using the hierarchy of coupled cluster methods in the basis set limit. First we compare the performance of AFQMC and CC methods for calculating absolute ground state energies in a small basis set where near-exact energies can be obtained. Fig. \ref{fig:heat} shows errors in ground state energies calculated using various methods with the cc-pVDZ basis and the geometries provided in Ref. \citenum{tajti2004heat}. We use CCSDTQP energies from Ref. \citenum{bomble2005coupled} as the near-exact reference. Ref. \citenum{bomble2005coupled} does not report the CCSDTQP energy for \ce{CO2}, for which we performed the calculation ourselves. For all 26 molecules in the dataset, AFQMC/CISD errors are lower than those from CCSD(T). The root mean square deviations (RMSD) are shown in the legend of Fig. \ref{fig:heat}. The ground states of these molecules are largely single-reference, therefore, CCSD(T) has a relatively small RMSD of 1.7 mH. AFQMC/CISD has an even smaller RMSD of 0.8 mH, with CCSDTQ energies being nearly exact. AFQMC/HF, as reported in Ref. \citenum{sukurma2023benchmark}, shows relatively small errors for most molecules but has some notable outliers. We confirmed that we obtained AFQMC/HF energies within error bars of those reported in Ref. \citenum{sukurma2023benchmark} for \ce{O2} and \ce{CN} molecules using our code. Interestingly, we find AFQMC/CISD energies to be variational for all of these molecules, whereas AFQMC/HF energies are not.

\begin{figure*}
   \centering
   \includegraphics[width=0.9\textwidth]{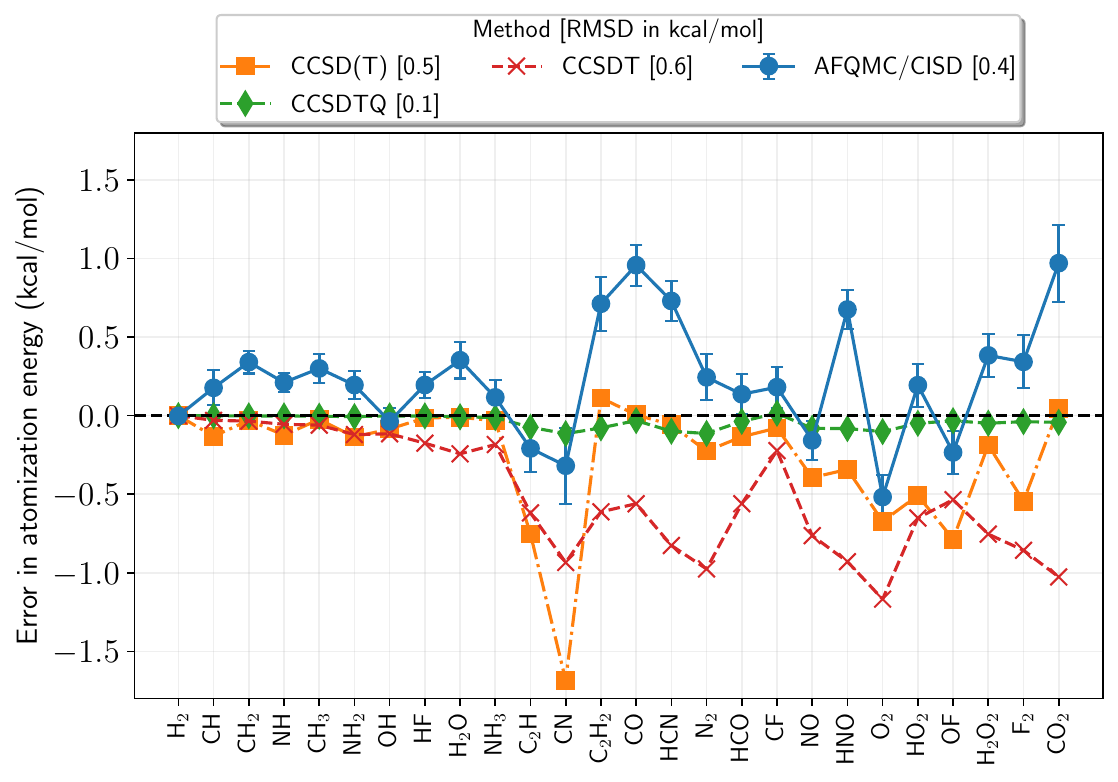}
   \caption{Errors in the valence correlation contribution to the atomization energy of molecules in the HEAT dataset in the basis set limit (extrapolated from cc-pVTZ and cc-pVQZ values). For CCSDTQ, the quadruples correction to the extrapolated CCSDT energy was evaluated in the cc-pVDZ basis. CCSDT and CCSDTQ energies are from Ref. \citenum{tajti2004heat}.}
   \label{fig:heat_atomization}
\end{figure*}
In lieu of calculating exact ground state energies for larger basis sets, we use a composite scheme as outlined in Ref. \citenum{tajti2004heat}, to obtain benchmark correlation contributions to atomization energies in the basis set limit. This analysis focuses solely on the correlation energy of the valence electrons. Using the frozen-core approximation, we calculate reference correlation energies for all molecules in the HEAT set and their constituent atoms as 
 \begin{equation}
   \begin{split}
      E_{\text{corr}}\left[\text{Ref}\right] &= E^{\text{TQ}}_{\text{corr}}\left[\text{CCSD(T)}\right] + \Delta E^{\text{TQ}}\left[{\text{CCSDT}}\right]\\ 
      &+ \Delta E^{\text{DZ}}\left[\text{CCSDTQ}\right] + \Delta E^{\text{DZ}}\left[\text{CCSDTQP}\right],
   \end{split}
\end{equation}
where \(E^{\text{TQ}}_{\text{corr}}\left[\text{CCSD(T)}\right]\) is the CCSD(T) correlation energy extrapolated to the continuum limit using cc-pVTZ and cc-pVQZ basis sets. \(\Delta E^{\text{TQ}}\left[{\text{CCSDT}}\right]\) is the iterative triples correction to the CCSD(T) value, also evaluated using the same two-point extrapolation approach. The \(\Delta E^{\text{DZ}}\left[\text{CCSDTQ}\right]\) and \(\Delta E^{\text{DZ}}\left[\text{CCSDTQP}\right]\) corrections are computed in the cc-pVDZ basis. The CCSDT and CCSDTQ corrections are taken from Ref. \citenum{tajti2004heat} and the CCSDTQP corrections from Ref. \citenum{bomble2005coupled}. While these correlation energies may not be near-exact in an absolute sense, they should provide a highly accurate estimate of the valence correlation contribution to atomization energies due to the cancellation of errors in the energy differences. We follow Ref. \citenum{tajti2004heat} in using UHF-based CCSD(T) for open-shell molecules for atomization energy calculations. We calculate the AFQMC/CISD correlation contributions using the two-point TZ-QZ extrapolation method. 

Fig. \ref{fig:heat_atomization} shows the errors in these contributions to atomization energies. We note that CCSD(T), CCSDT, and AFQMC/CISD values are all calculated in the basis set limit, while the CCSDTQ values include the quadruples correction only in the cc-pVDZ basis. As noted in many previous studies, despite significant errors in ground state energies, CCSD(T) benefits from near-perfect error cancellation between molecular and atomic energies, resulting in small errors in the atomization energy with an RMSD of just 0.5 kcal/mol, primarily due to the outlier \ce{CN}. This performance is, in most cases, superior to the more expensive CCSDT method, which has a slightly higher RMSD of 0.6 kcal/mol. While AFQMC/CISD does not benefit from error cancellation to the same extent as CCSD(T), it still achieves a lower overall RMSD of 0.43(2) kcal/mol. The AFQMC/CISD errors are smaller than 1 kcal/mol for all molecules, with no extreme outliers. CCSDTQ values are almost exact in all cases with an RMSD of 0.1 kcal/mol, consistent with the high accuracy of this method for absolute ground state energies. 

It is worth mentioning that AFQMC/CISD does not suffer from the large errors in atomic energies that are observed in AFQMC/HF, as reported in Ref. \citenum{lee2022twenty}. To illustrate this point, we present a comparison of the errors in atomic energies calculated using different methods in Appendix \ref{app:atomic}.

\begin{figure*}[t]
   \centering
   \includegraphics[width=0.8\textwidth]{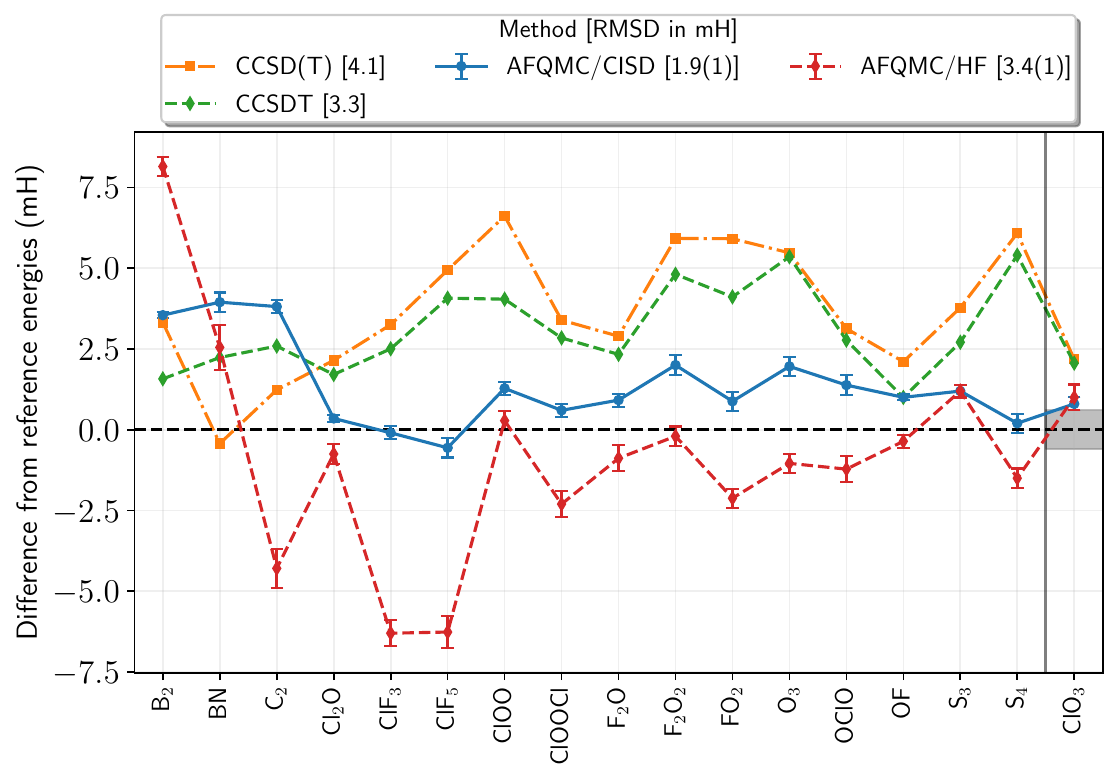}
   \caption{Errors in ground state energies for molecules in the W4-MR dataset (cc-pVDZ basis) calculated using various methods. The reference energies for all molecules except \ce{ClO3} (on the left of the vertical line) are CCSDTQ and for \ce{ClO3} it is AFQMC/HCI converged with respect to the number of determinants, with the gray shaded area showing the stochastic error estimate.}
   \label{fig:w4mr}
\end{figure*}
\subsubsection{W4-multireference dataset}\label{sec:w4mr}
The W4 dataset, created by Martin and co-workers, has also been widely used for assessing the quality of new electronic structure methods.\cite{karton2017w4} Here, we focus on the subset of molecules with significant multireference character grouped in the W4-MR set. These molecules were identified using a diagnostic based on the contribution of the perturbative triples to the correlation energy. For these challenging cases, CCSD(T) is known to show larger errors. To determine whether AFQMC based on a CISD trial can outperform CCSD(T), we compare ground state energies using a small basis set. Fig. \ref{fig:w4mr} shows the difference in ground state energies for various methods using CCSDTQ energies as reference. We used geometries provided in Ref. \citenum{karton2017w4} and the cc-pVDZ basis set. Since these molecules are larger than those in the HEAT set, it is computationally expensive to perform CCSDTQP calculations for all of them, therefore we use CCSDTQ as reference, which, although not exact, should nonetheless provide an accurate reference point. To confirm the accuracy of CCSDTQ energies, we performed AFQMC calculations with HCI trial states for the five largest molecules in this set viz. \ce{S4}, \ce{ClOOCl} \ce{ClF5}, \ce{ClO3}, and \ce{F2O2}. We can converge the AFQMC energies with respect to the number of determinants in the HCI expansion, to obtain near exact energies. In all these molecules, except \ce{ClO3}, we found CCSDTQ to be within 1 mH of AFQMC/HCI energies. We could not converge CCSDTQ for the open-shell \ce{ClO3} molecule with the available computational resources, therefore we used a converged AFQMC/HCI energy as the reference value for this molecule. 

From Fig. \ref{fig:w4mr} we see that CCSD(T) errors are substantially larger compared to the HEAT set, with an RMSD of 4.2 mH. A fully iterative treatment of the triples in CCSDT improves the accuracy only slightly, reducing the RMSD to 3.4 mH. Interestingly, the more economical AFQMC/HF method achieves the same RMSD as CCSDT for this set of molecules. AFQMC/CISD, on the other hand, exhibits differences of less than 2 mH from CCSDTQ in most cases, with an RMSD of 2.0(1) mH. Most of this error is attributed to three molecules: \ce{BN}, \ce{B2}, and \ce{C2}. \ce{BN} and \ce{C2} are isoelectronic and are both known to be pathological cases where even explicitly multireference approaches struggle to describe the correlation.\cite{watts1992coupled,martin1992ab} Note that CCSD(T) energies are lower than CCSDT in both these molecules. It is interesting to observe that AFQMC/CISD energies in many cases are between CCSD(T) and AFQMC/HF energies, where the former tend to be undercorrelated and the latter overcorrelated. We see this pattern repeated in other results reported in this paper. Overall, AFQMC/CISD shows significantly smaller errors than CCSD(T) for these multireference molecules.

\begin{figure*}[t]
   \centering
   \includegraphics[width=0.8\textwidth]{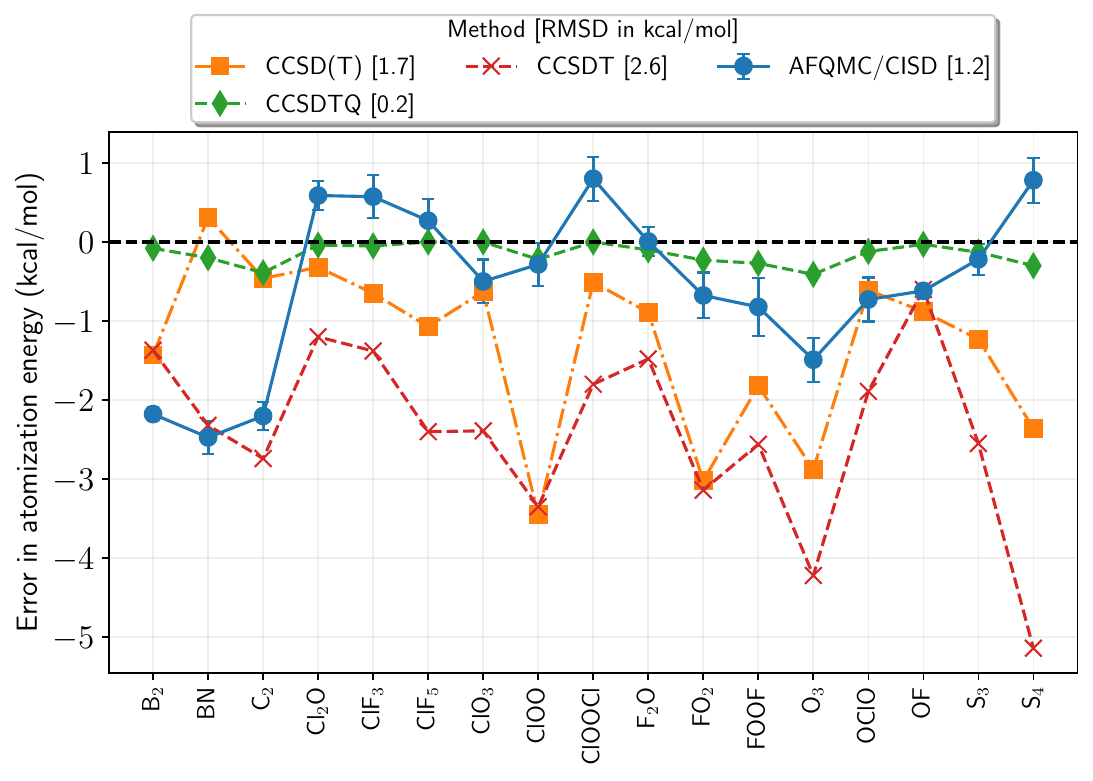}
   \caption{Errors in the valence correlation contribution to the atomization energy (extrapolated from cc-pVDZ and cc-pVTZ values) of molecules in the W4-MR dataset. For CCSDTQ, the quadruples correction to the CCSDT energy was evaluated in the cc-pVDZ basis. CCSDT and CCSDTQ energies are from Ref. \citenum{karton2017w4}.}
   \label{fig:w4mr_atomization}
\end{figure*}

We also calculate the valence correlation contributions to the atomization energies of the W4-MR molecules. Due to their significant multireference character, higher-order CC corrections are larger for these molecules than for the HEAT set. In Ref. \citenum{karton2017w4}, the full triples corrections are estimated using extrapolations from cc-pVDZ and cc-pVTZ energies, while higher-order corrections are evaluated in even smaller basis sets. To enable a fair comparison, we employ a DZ-TZ extrapolation for the AFQMC/CISD correlation contributions. Although this extrapolation does not provide results very close to the basis set limit, it allows us to compare the accuracy of different methods in capturing correlation while minimizing the influence of basis set effects. Thus, we use the following composite scheme to obtain the reference valence correlation energies:

\begin{equation}
   \begin{split}
      E_{\text{corr}}\left[\text{Ref}\right] &= E^{\text{DT}}_{\text{corr}}\left[\text{CCSD(T)}\right] + \Delta E^{\text{DT}}\left[{\text{CCSDT}}\right]\\ 
      &+ \Delta E^{\text{DZ}}\left[\text{CCSDTQ}\right] + \Delta E\left[\text{higher order}\right],
   \end{split}
\end{equation}
where corrections beyond perturbative triples are taken from Ref. \citenum{karton2017w4}. 

Fig. \ref{fig:w4mr_atomization} shows the errors in these contributions to atomization energies. CCSD(T) exhibits larger errors in atomization energies compared to the HEAT set, with an RMSD of 1.7 kcal/mol, which is consistent with the greater errors observed in the ground state energies of these multireference molecules. CCSDT has a significantly larger RMSD of 2.6 kcal/mol, despite showing smaller errors in ground state energies, again highlighting the benefit of error cancellation in CCSD(T). AFQMC/CISD is more accurate than CCSD(T) in nearly all cases, with a smaller RMSD of 1.17(5) kcal/mol. This value is dominated by the outliers \ce{BN}, \ce{B2}, and \ce{C2}, where AFQMC/CISD errors exceed 2 kcal/mol. CCSDTQ energies are almost exact in all cases, with an RMSD of just 0.2 kcal/mol. 

\begin{figure*}[t]
  \centering
  \includegraphics[width=0.75\textwidth]{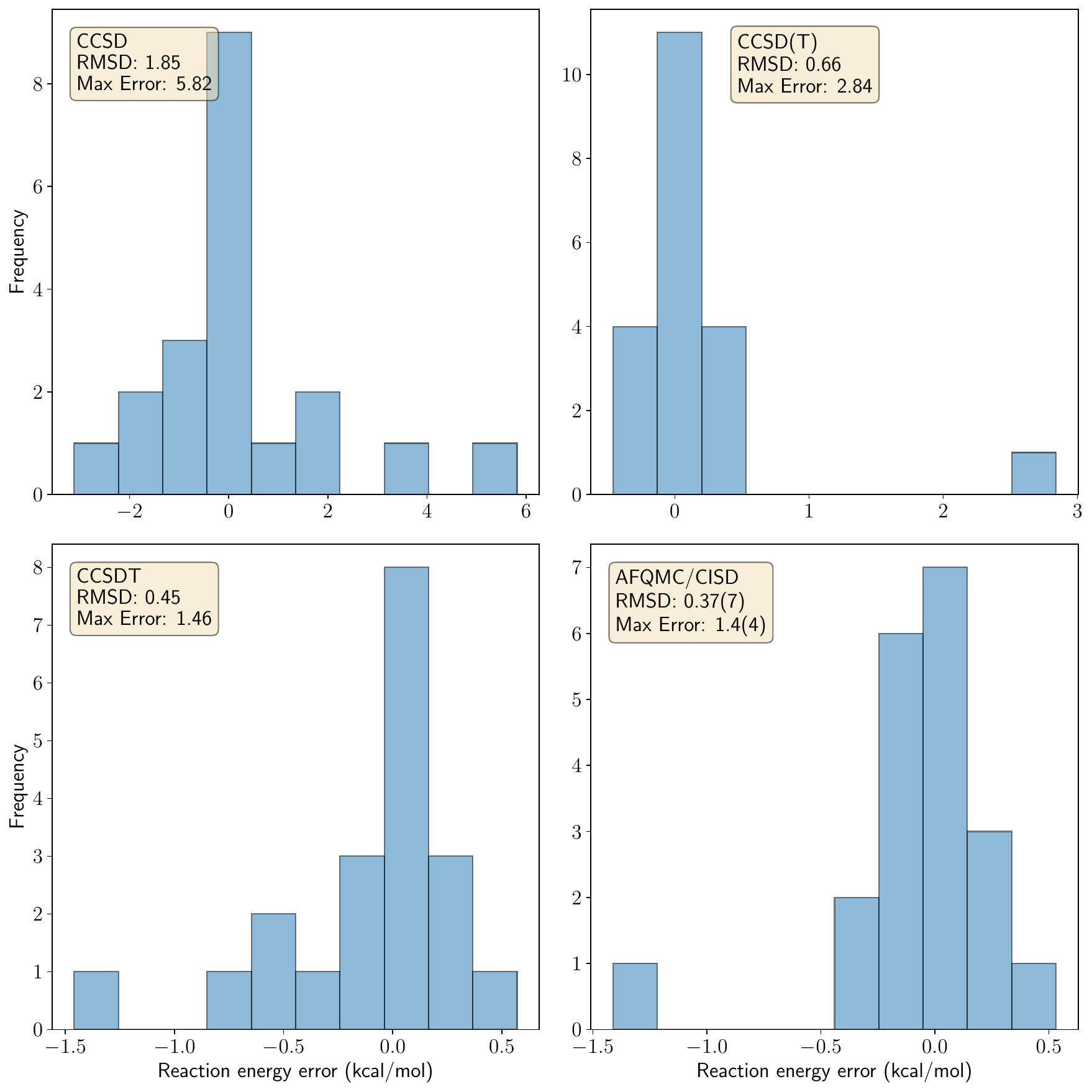}
  \caption{Distribution of errors in the valence correlation contribution (extrapolated from cc-pVDZ and cc-pVTZ values) to isomerization energies for the ISO20 subset of W4\cite{karton2017w4} using different methods.}
  \label{fig:iso}
\end{figure*}
\subsubsection{Isomerization energies}\label{sec:iso}
While the calculation of isomerization energies is not as difficult for electronic structure methods as the calculation of atomization energies, they are arguably more useful for practical applications. Even though the electronic structure can change drastically during isomerization, these calculations often benefit more from error cancellations. In this section, we first consider the ISO20 subset of the W4 dataset, which contains 20 isomerization reactions of small molecules. We calculate the reference valence correlation contributions to isomerization energies using the same composite scheme as for the atomization energies used in section \ref{sec:w4mr}. Again, we use the reference values reported in Ref. \citenum{karton2017w4}. Fig. \ref{fig:iso} shows the distribution of errors in the isomerization energies for various methods. Of the presented methods, CCSD exhibits the worst performance with an RMSD of 1.85 kcal/mol. On the other hand, we find AFQMC/CISD to be the most accurate of the presented methods with an RMSD of 0.37(7) kcal/mol. CCSD(T) has a larger RMSD of 0.66 kcal/mol, where the worse performance is mostly attributable to one outlier, \ce{O2Cl}\(\rightarrow\)\ce{OOCl}, with both these molecules belonging to the MR subset. In contrast to the atomization energies, CCSDT is more accurate than CCSD(T) in this case, with an RMSD of 0.45 kcal/mol.

\begin{table}
  \caption{Automerization barrier of cyclobutadiene. The first two columns show absolute energies in Hartree for the minimum and transition state geometries, and the last column shows the barrier height \(E_{\text{TS}} - E_{\text{min}}\) in kcal/mol. The free projection (fp)-AFQMC values are the best theoretical estimates.}\label{tab:cbd}
  \centering
  \begin{tabular}{*{7}c}
  \hline
  &~& Min &~& TS &~& $\Delta E$ (kcal/mol) \\
  \hline
  CCSD(T) && -154.3906 && -154.3615 && 18.3 \\
  UCCSD(T) && -154.3842 && -154.3731 && 6.7 \\
  AFQMC/HF && -154.3961(8) && -154.3769(7) && 12.0(7) \\
  AFQMC/CISD && -154.3928(5) && -154.3768(3) && 10.0(4) \\
  fp-AFQMC\cite{mahajan2021taming} && -154.3929(5) && -154.3767(5) && 10.2(4) \\
  \hline
  \end{tabular}
\end{table}

We also consider the automerization barrier of cyclobutadiene, which is challenging due to the multireference character of the transition state. The minimum energy structure has a \(D_{2h}\) symmetry, while the transition state has a \(D_{4h}\) structure. The higher symmetry of the latter leads to a substantial biradical character. We use the near-exact free-projection (fp) AFQMC energies reported in Ref. \citenum{mahajan2021taming} as benchmark values for comparison. We used the same geometries as in that study and the cc-pVTZ basis set. Table \ref{tab:cbd} shows the absolute energies and barriers obtained using different methods. CCSD(T) based on the RHF reference significantly undercorrelates the transition state leading to an overestimation of the barrier height by 8.1(4) kcal/mol, a well known failure of the perturbative triples correction in this system. The use of UHF reference in UCCSD(T) while fixing the error for the transition state to a large extent, leads to an underestimation of the gap by 3.5(4) kcal/mol. Both AFQMC/HF and AFQMC/CISD perform much better with errors of 1.8(8) and -0.2(6) kcal/mol, respectively. AFQMC/CISD, in particular, agrees with even the absolute fp-AFQMC energies within stochastic error bars.

\begin{figure}[t]
   \centering
   \includegraphics[width=0.45\textwidth]{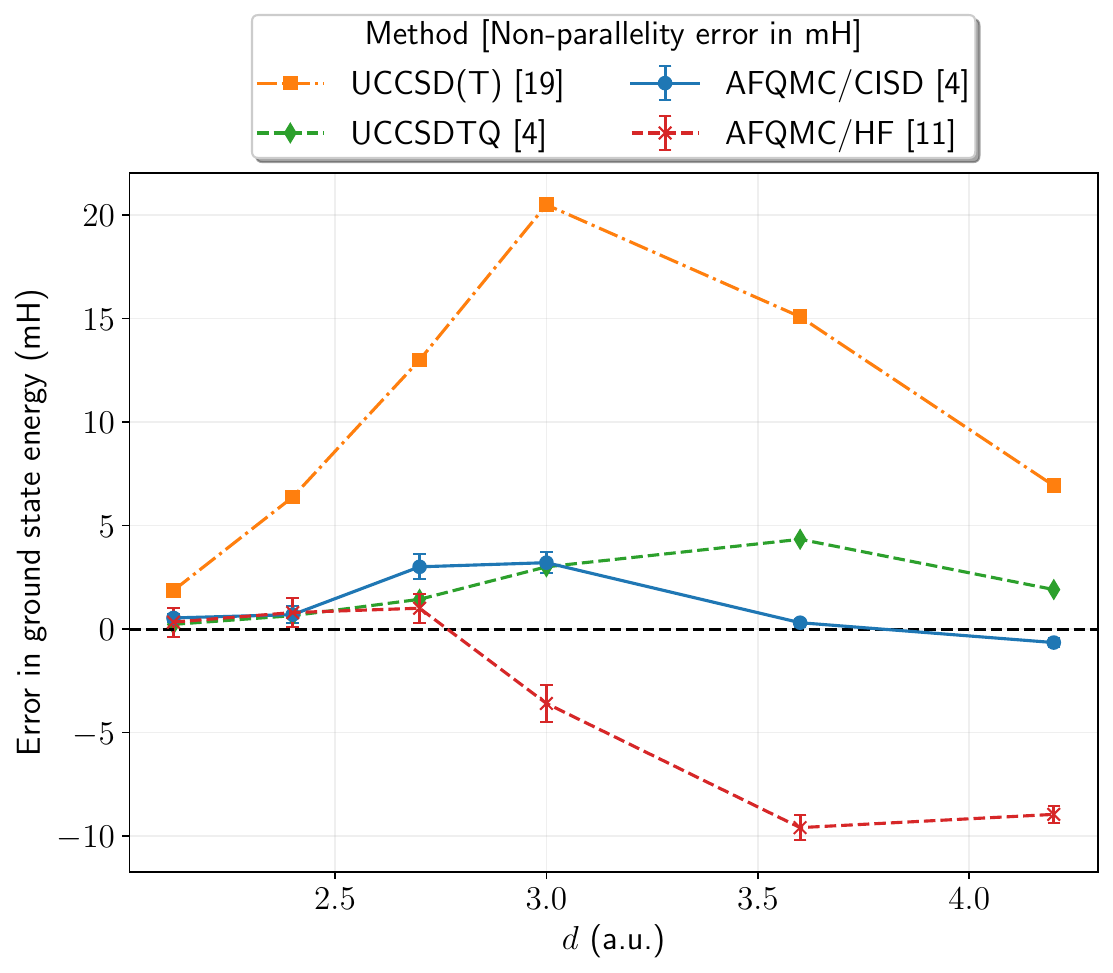}
   \caption{Errors in the ground state energy of \ce{N2} at different bond lengths (cc-pVDZ basis). UHF-based reference states were used in all methods. Exact and CC energies are taken from Ref. \citenum{chan2004state}.}
   \label{fig:n2}
\end{figure}

\subsubsection{\texorpdfstring{\ce{N2}}{N2} dissociation}
Bond breaking is ubiquitous in various chemical processes and poses a significant challenge for many electronic structure methods due to the changing nature of correlation at different geometries. In this context, we calculate the potential energy curve of the \ce{N2} molecule, which is particularly challenging due to its triple bond. Fig. \ref{fig:n2} shows the errors in the ground state energy at different bond lengths using the cc-pVDZ basis set. Exact energies are taken from Ref. \citenum{chan2004state}, which also provides high-order CC energies. We present CC energies based on the UHF reference here, as RHF-based energies can diverge at stretched geometries. The non-parallelity error, defined as the difference between the maximum and minimum errors in absolute energies, is a useful metric for identifying methods that offer a balanced treatment of electronic correlation across the potential energy curve. The perturbative treatment of triples in UCCSD(T) results in a large non-parallelity error of 19 mH due to massive undercorrelation in the intermediate stretched region. In contrast, AFQMC/CISD has a non-parallelity error of 4 mH, significantly smaller than that of UCCSD(T), and comparable to UCCSDTQ. AFQMC/HF performs better than UCCSD(T) but still shows a substantial non-parallelity error of 11 mH. 

\subsubsection{Dipole moments}
\begin{table*}
   \caption{Dipole moments (in a.u.) of \ce{CO} and \ce{CH2O} molecules at equilibrium geometries using the aug-cc-pVQZ basis set.}\label{tab:dipoles}
   \centering
   \begin{tabular}{*{13}c}
   \hline
   &~~& \ce{H2O} &~~& \ce{NH3} &~~& \ce{HCl} &~~& \ce{HBr} &~~& \ce{CO} &~~& \ce{CH2O} \\
   \hline
   HF && 0.7810 && 0.6364 && 0.4680 && 0.3726 && -0.1147 && 1.128 \\
   CCSD && 0.7369 && 0.6047 && 0.4370 && 0.3392 && 0.0156 && 0.9738 \\
   AFQMC/HF && 0.714(2) && 0.582(6) && 0.427(3) && 0.325(3) && 0.010(3) && 0.986(4) \\
   CCSD(T) && 0.7281 && 0.5969 && 0.4325 && 0.3343 && 0.0393 && 0.9454 \\
   CCSDT && - && - && - && - && 0.0370 && 0.9465 \\
   AFQMC/CISD && 0.726(2) && 0.589(3) && 0.428(2) && 0.331(2) && 0.035(2) && 0.932(3) \\
   \hline
   Experiment && 0.730(2)\cite{lide2004crc} && 0.581(1)\cite{shimizu1970stark} && 0.430\cite{lovas2005diatomic} && 0.325\cite{lovas2005diatomic} && 0.048(1)\cite{meerts1977electric} && 0.918(1)\cite{theule2003fluorescence} \\
   \hline
   \end{tabular}
\end{table*}

In a recent paper,\cite{mahajan2023response} we presented a method to calculate properties other than energy in AFQMC within a response formalism, leveraging automatic differentiation. This approach provides superior statistical estimators and better computational efficiency than the back-propagation technique. Since our AFQMC/CISD code is developed within this differentiable framework, we can straightforwardly calculate first-order properties using this new trial state. We refer the reader to the original paper\cite{mahajan2023response} for details of the methodology and leave a more thorough investigation of such calculations for future work. Here, we briefly discuss the calculation of dipole moments of some of the same molecules, including the two challenging cases identified previously, viz. \ce{CO} and \ce{CH2O}. Table \ref{tab:dipoles} presents the results of these calculations using the aug-cc-pVQZ basis set and the same geometries as in Ref. \citenum{mahajan2023response}. We believe this basis set is sufficiently large for meaningful comparison with experimental measurements. The CC and AFQMC/HF dipole moments include contributions from HF orbital response. The AFQMC/CISD values do not include trial response contributions. We observe that CCSD and AFQMC/HF both show similar errors with respect to experimental values. These errors are substantial for the two challenging cases, particularly for \ce{CO}, where HF predicts an incorrect polarity direction. We therefore also performed CCSDT calculations for \ce{CO} and \ce{CH2O}. CCSD(T), CCSDT, and AFQMC/CISD show much better agreement with the experimental values. One should note, however, that aside from comparison with experimental values, which are complicated by experimental uncertainties, the consistency between AFQMC/CISD and high-order CC dipole moments provides confidence in their accuracy. Determining the relative accuracy of these two methods for property evaluation requires further careful investigation.

\subsection{Transition metal containing molecules}\label{sec:transition_metal}
In this section, we examine molecules and clusters containing 3\(d\) transition metal elements, which pose a more challenging problem for electronic structure methods. In general, CC methods do not achieve the same level of accuracy for these systems as they do for the lighter main group molecules. 

\begin{figure}[t]
   \centering
   \includegraphics[width=0.45\textwidth]{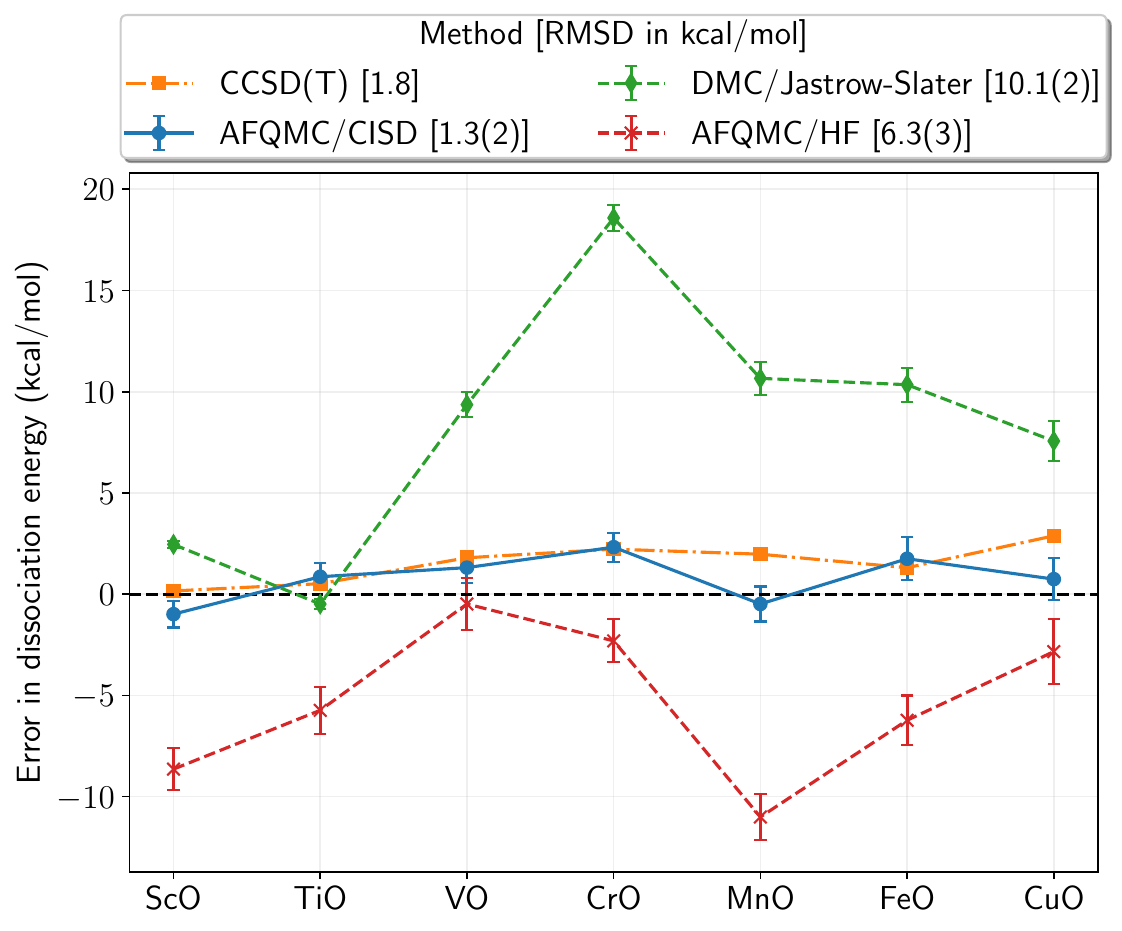}
   \caption{Errors in the dissociation energies of transition metal oxide molecules. We performed extrapolations to the basis set limit based on energies obtained using the TZ and QZ Trail-Needs pseudopotential and basis sets. The errors are with respect to near-exact SHCI values taken from Ref. \citenum{williams2020direct}, which is also the source of DMC energies.}
   \label{fig:tmo}
\end{figure}
\subsubsection{Transition metal binary compounds}
Fig. \ref{fig:tmo} shows the errors in dissociation energies of seven transition metal oxide molecules obtained using CC and QMC approaches. We use the near-exact SHCI energies reported in Ref. \citenum{williams2020direct} for these molecules in large basis sets as reference values for our analysis. To allow a direct comparison, we follow this work in using the same bond lengths and the Trail-Needs pseudopotential and corresponding basis sets. For all orbital space methods, we performed an exponential fit for HF energies in the TZ, QZ, and 5Z basis sets, and an inverse cubic fit for correlation energies in the TZ and QZ basis sets to extrapolate the total energies to the basis set limit. Since diffusion Monte Carlo (DMC) inherently works directly in the continuum limit, we used the dissociation energies directly from Ref. \citenum{williams2020direct}. We again find AFQMC/CISD to be more accurate than CCSD(T) with an RMSD of 1.3(2) kcal/mol. Both methods benefit from some cancellation of errors, as can be seen from the absolute energies reported in the supporting information.\cite{afqmc_files} Both DMC and AFQMC/HF are lower scaling approaches based on a single determinant trial state (a Jastrow factor is used in DMC), but AFQMC/HF has a smaller RMSD of 6.3(3) kcal/mol. We note that DMC energies can be improved by adding more determinants in the trial state as well as by including backflow correlations. But it is also worth mentioning that with the commonly used Jastrow-Slater trial state, DMC can exhibit substantial errors, and one should be cautious in using it as a benchmark method for transition metal systems. 

\begin{figure*}
  \centering
  \subfloat[]{
  \includegraphics[width=0.9\columnwidth]{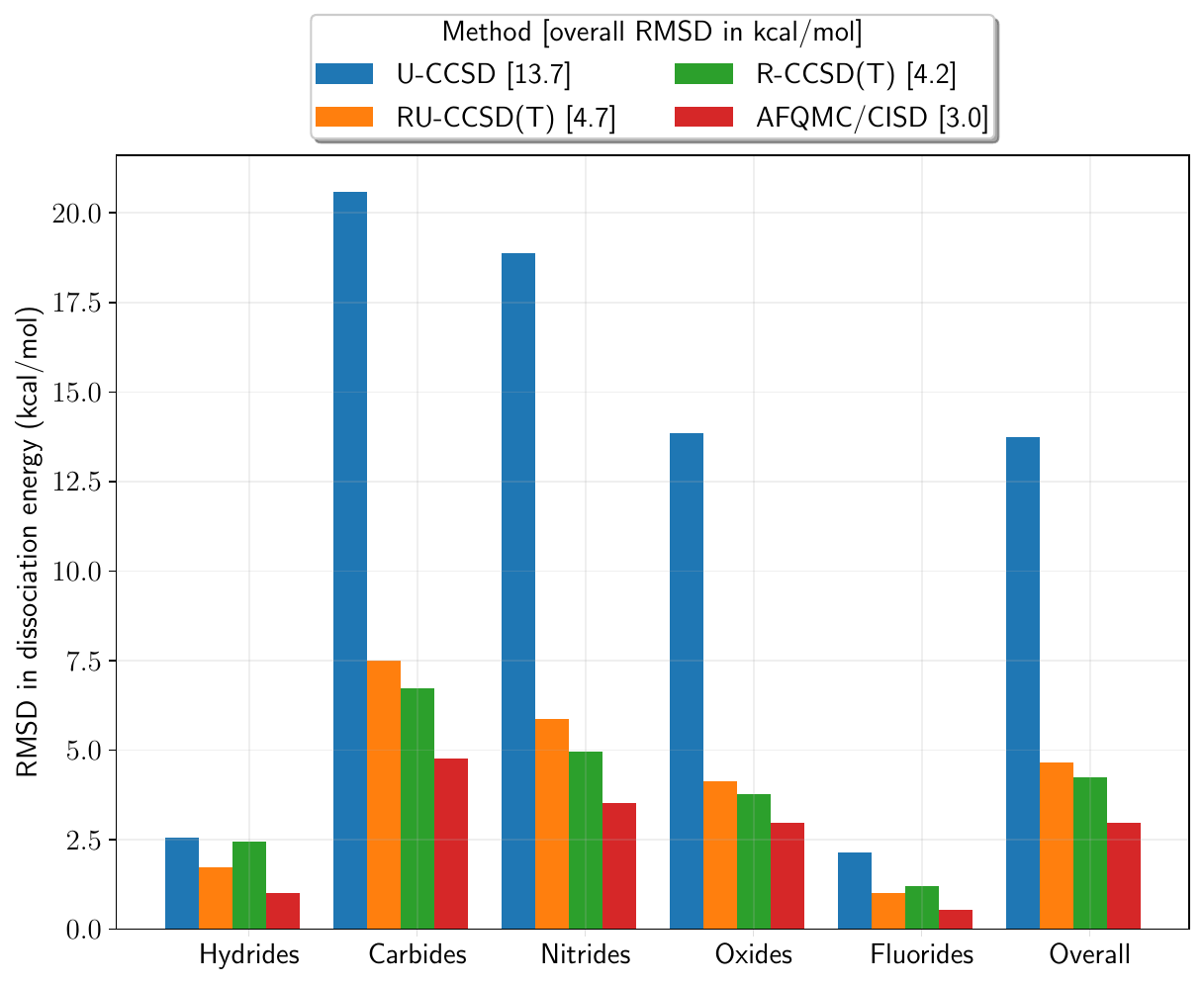}
  }\qquad
  \subfloat[]{
     \includegraphics[width=0.9\columnwidth]{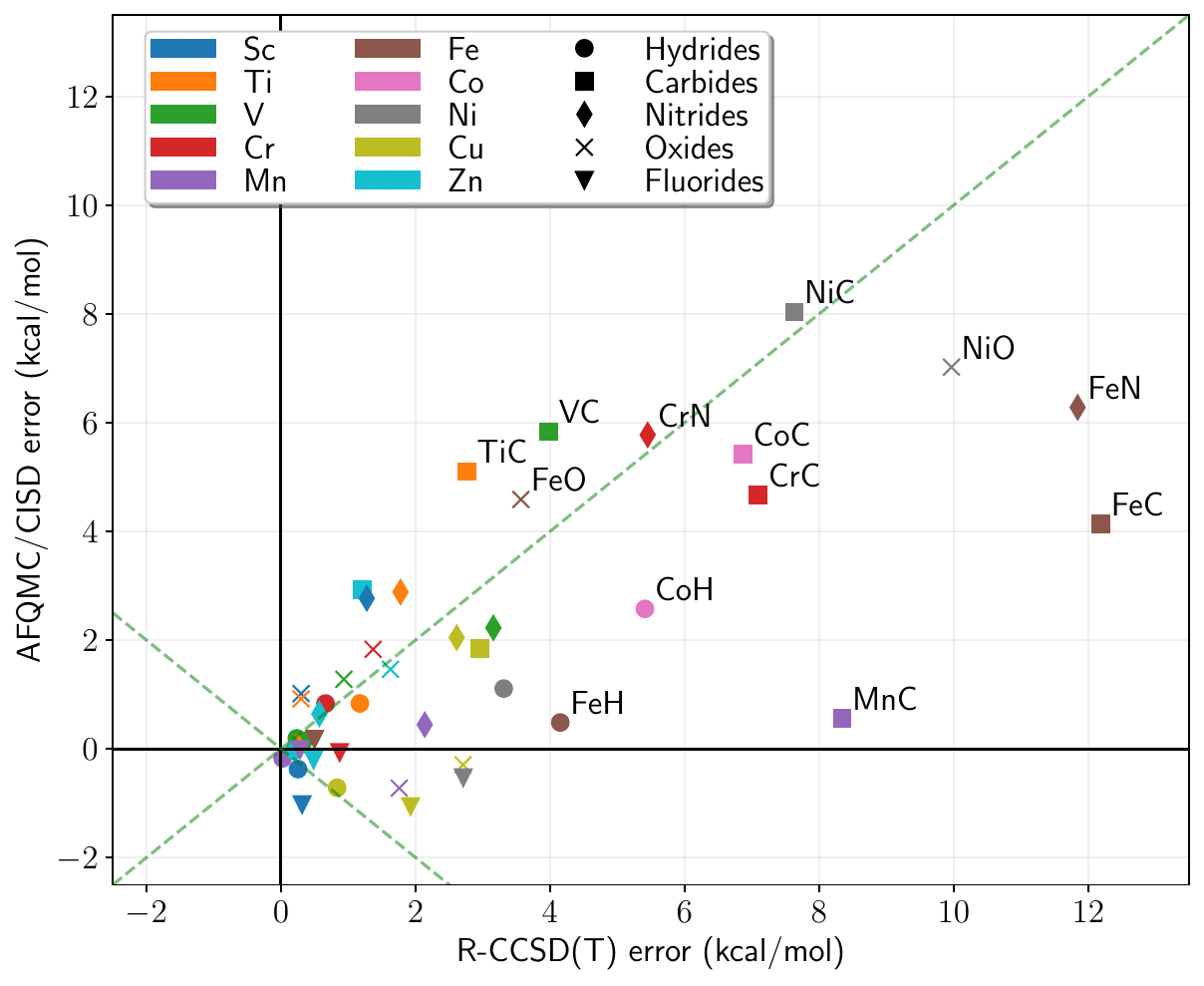}
  }
  \caption{Errors in the atomization energies of transition metal binary compounds in the def2-SVP basis compared to the near-exact selected CI values reported in Ref. \citenum{hait2019levels} (a) RMSD values for various methods in different categories. RU-CCSD(T) refers to the protocol of using the RHF reference for closed-shell species and UHF reference for open-shell ones. (b) Scatter plot showing a direct comparison of the performance of R-CCSD(T) and AFQMC/CISD for all compounds.}\label{fig:tm_dimers}
\end{figure*}

We also calculated atomization energies of \(3d\) transition metal binary compounds, including hydrides, carbides, nitrides, oxides, and fluorides, in the double-\(\zeta\) def2-SVP basis set (with some exceptions\cite{tm_dimer_exceptions}). We use the near-exact energies reported in Ref. \citenum{hait2019levels} using a selected CI method as a reference. Fig. \ref{fig:tm_dimers} shows the errors in the atomization energies calculated using AFQMC/CISD and different CC protocols. While CCSD and higher order methods are relatively insensitive to the choice of reference determinant orbitals in most cases, it can make a nontrivial difference for transition metal systems. We tested three different strategies: restricted reference determinants for all species (R-CC), unrestricted reference determinants for all species (U-CC), restricted reference determinants for closed-shell species, and unrestricted reference determinants for open-shell species (RU-CC). We found R-CCSD(T) to be the most accurate overall with an RMSD of 4.2 kcal/mol, while U-CCSD(T) was the least accurate with an RMSD of 5.2 kcal/mol. RU-CCSD(T) performs better than R-CCSD(T) for hydrides and fluorides but is worse for other compounds and worse overall with an RMSD of 4.7 kcal/mol (see left panel of Fig. \ref{fig:tm_dimers}). AFQMC/CISD is the best performing of the presented methods for all categories and overall as well with an RMSD of 3.0 kcal/mol. We also show the errors in U-CCSD which is used to generate the trials for AFQMC, and except for hydrides, it exhibits much larger errors than all other methods. The right-hand panel of Fig. \ref{fig:tm_dimers} shows a comparison of the performance of R-CCSD(T) and AFQMC/CISD for all compounds. It is evident that AFQMC/CISD is more accurate than R-CCSD(T) for most molecules with the largest errors for both methods being for carbides and nitrides. 

\subsubsection{\texorpdfstring{\ce{[Cu2O2]^2+}}{[Cu2O2]2+} Isomerization}
\begin{table}
   \caption{Isomerization energy of \ce{[Cu2O2]^2+}. The first two columns show absolute energies in Hartree for the two geometries, and the last column shows the isomerization energy (\(E(f=0) - E(f=1)\)) in kcal/mol. The free projection (fp)-AFQMC values are the best theoretical estimates.}\label{tab:cu2o2}
   \centering
   \begin{tabular}{*{7}c}
   \hline
   &~& \(f=0\) &~& \(f=1\) &~& $\Delta E$ (kcal/mol) \\
   \hline
   CCSD(T) && -542.0885 && -542.1373 && 30.6 \\
   AFQMC/HF && -542.0966(7) && -542.152(1) && 34.8(8) \\
   AFQMC/CISD && -542.0906(9) && -542.1290(9) && 24.1(8) \\
   fp-AFQMC\cite{mahajan2021taming} && -542.0964(7) && -542.1348(7) && 24.1(6) \\
   \hline
   \end{tabular}
\end{table}
The isomerization pathway between two structures of the \ce{[Cu2O2]^2+}, an active site in enzymes like tyrosinase and catechol,\cite{solomon1992electronic} has been studied using numerous computational methods.\cite{cramer2006theoretical} We recently reported benchmark near-exact free projection (fp)-AFQMC energies for this system, demonstrating that methods previously considered highly accurate for this system deviate from these reference values by as much as 10 kcal/mol.\cite{mahajan2021taming} An AFQMC/HCI calculation that systematically converged the phaseless error with respect to the number of determinants in the trial found energies in good agreement with the free projection results.\cite{malone2022ipie} Table \ref{tab:cu2o2} shows absolute energies (in Hartrees) and isomerization energies (in kcal/mol) calculated using different methods along with the reference fp-AFQMC values. We used the same basis set and frozen core as that employed in Ref. \citenum{mahajan2021taming} (referred to as BS1 in that work), which consists of a Stuttgart pseudopotential and associated basis functions on the copper atoms and an ANO triple-zeta basis set on the oxygens. The
semicore \(3s\) and \(3p\) electrons on copper and the core 1s electrons on oxygen were frozen at the HF level leading to a correlation space (32e, 108o). CCSD(T) does not accurately describe the energetics of this system, overestimating the isomerization energy by 6.5(6) kcal/mol. AFQMC/HF performs even worse, with an error in the gap of about 10(1) kcal/mol, primarily due to substantial overcorrelation of the \(f=1\) isomer. In contrast, the AFQMC/CISD gap is in remarkable agreement with fp-AFQMC results, within the stochastic error margin seemingly benefiting from a cancellation of errors in the absolute energies of the two isomers. This system illustrates the accuracy of AFQMC/CISD in a case where CCSD(T) fails significantly.

\subsubsection{Singlet-Quintet gap in \texorpdfstring{\ce{[Fe(H2O)6]^2+}}{[Fe(H2O)6]2+}}

\begin{table*}[htp]
   \caption{Singlet(LS) - Quintet(HS) splitting \(\Delta E = E_{\text{HS}} - E_{\text{LS}}\) of \ce{[Fe(H2O)6]^2+} in kcal/mol.}\label{tab:fehydro}
   \centering
   \begin{threeparttable}
   \begin{tabular}{ccccccc}
   \hline
   \hline
   Basis set &~~& Method &~~& Correlation space &~~& \(\Delta E\)\\
   \hline
   \ce{Fe}: cc-pwCVTZ-DK && AFQMC/HF && (62e, 235o) && -44.2(8) \\
   \ce{O}, \ce{H}: cc-pVDZ-DK && AFQMC/CISD  &&  && -36.1(9) \\
   && CCSD(T) &&  && -37.7 \\
   && DLPNO-CCSD(T1) &&  && -36.7 \\
   \hline
   \ce{Fe}: cc-pwCVTZ-DK && AFQMC/HF && (62e, 439o) && -46(1) \\
   \ce{O}, \ce{H}: cc-pVTZ-DK && AFQMC/CISD  &&  && -36(1) \\
   && DLPNO-CCSD(T1)\cite{flöser2020detailed} &&  && -36.5 \\
   \hline
   ANO-RCC triple-\(\zeta\) && CASPT2\cite{pierloot2006relative} && (62e, 301o) && -46.6\\
   \hline
   Continuum limit && DMC/Jastrow-Slater\cite{song2018benchmarks} && 62e && -41.0\\
   && AFQMC/CISD && 62e, MP2 CBS correction && -32(1)\\
   && DLPNO-CCSD(T1)\cite{flöser2020detailed} && 62e, Q-5Z extrapolation && -33.3\\
   \hline
   \hline
   \end{tabular}
   \end{threeparttable}
\end{table*}

Some transition metal complexes can undergo a change in their spin state in response to a small perturbation, such as a change in geometry -- a phenomenon known as spin crossover. Calculating the spin-state splitting in these systems is a difficult task due to the distinct nature of the two electronic states: one being a high spin state, the other a low spin state. Here we consider a system that exhibits this behavior: the \ce{[Fe(H2O)6]^2+} octahedral complex. Table \ref{tab:fehydro} presents the spin-state splitting \(\Delta E\) calculated using AFQMC, CC, CASPT2, and DMC methods. We use the geometries reported in Ref. \citenum{flöser2020detailed}, which investigated spin crossover in this and two other Fe(II) complexes using the domain-based pair natural orbital coupled cluster with singles, doubles, and perturbative triples (DLPNO-CCSD(T)) method.\cite{riplinger2013efficient} Since no near-exact results are available for this system, we compare our results with a variety of approximate methods that have been used and attempt to identify correlations. This assessment is further complicated by the use of different basis sets and geometries across different studies.

The most straightforward comparison between AFQMC and CCSD(T) can be made in a smaller basis set where both methods are applicable. For this comparison, a triple-zeta basis set was used for the Fe atom, while a double-zeta basis was used for the H and O atoms. Scalar relativistic effects were included using the X2C Hamiltonian, whereas Ref. \citenum{flöser2020detailed} used the DKH2 Hamiltonian.  We do not anticipate this difference will affect the spin-state splitting significantly. Our results show that AFQMC/CISD and canonical CCSD(T) agree within 1.6(9) kcal/mol. Based on the benchmark results presented here, we expect AFQMC/CISD to provide a more accurate estimate. The DLPNO approximation yields a gap within one kcal/mol of the canonical CCSD(T) method in this case. However, AFQMC/HF results in a considerably larger gap due to overcorrelation of the high-spin state and undercorrelation of the low-spin state relative to AFQMC/CISD values (absolute energies are provided in the supplementary information). 

Using the TZ basis for the entire cluster leads to a correlation space of (62e, 439o), for which we could not converge CCSD(T) calculations due to disk space constraints. This is the largest AFQMC/CISD calculation reported in this work, involving \(\approx 0.5 \times 10^9\) determinants. To our knowledge, this represents the largest trial wavefunction used in any real-space or orbital-space projection QMC method. Each AFQMC/CISD calculation required about 25 hours on a single GPU. The DLPNO-CCSD(T1) gap reported in Ref. \citenum{flöser2020detailed} is in excellent agreement with AFQMC/CISD. This close agreement between the two methods increases confidence in their accuracy for this system. It is also worth noting that increasing the basis set size on the ligand atoms results in only a minor change in the spin-splitting. The TZ quality ANO basis set used in the CASPT2 study reported in Ref. \citenum{pierloot2006relative} has roughly the same number of basis functions on Fe but fewer on the ligands. Given the relative insensitivity of the gap to the ligand basis size, we believe that the CASPT2 gap is likely off by about 10 kcal/mol from the correct value. 

The DLPNO-CCSD(T1) calculations in larger basis sets show a significant change in the gap of about 3 kcal/mol when going from TZ to the QZ/5Z extrapolated value. It is challenging to perform canonical AFQMC/CISD calculations with more than about 600 orbitals on a single GPU due to memory limitations, and our current implementation does not parallelize across multiple GPUs. Therefore we employ the MP2 complete basis set (CBS) correction which was shown to be very accurate for spin-state splittings in Ref. \citenum{drabik2024approaching}. We correct the AFQMC/CISD gap in the TZ basis using the TZ-QZ extrapolated MP2 gap, as 
\begin{equation}
    \Delta E_{\text{AFQMC/CISD}}^{\text{CBS}} \approx \Delta E_{\text{AFQMC/CISD}}^{\text{TZ}} + \Delta E_{\text{MP2}}^{\text{CBS}} - \Delta E_{\text{MP2}}^{\text{TZ}},
\end{equation}
where all the energies refer to correlation energies. The MP2 CBS corrected AFQMC/CISD splitting of -32(1) kcal/mol is in excellent agreement with the QZ/5Z extrapolated DLPNO-CCSD(T1) value of -33.3. The close agreement between different accurate methods leads us to believe that this is the best estimate of the spin-state splitting in the continuum limit. Interestingly, the DMC/Jastrow-Slater estimate of the gap reported in Ref. \citenum{song2018benchmarks} differs from this value by about 9 kcal/mol, highlighting the challenges of using this method for benchmark calculations in transition metal systems. 

\subsection{Energy and computational cost scaling}\label{sec:cost}
In this section, we examine the asymptotic behavior of AFQMC/CISD in terms of ground-state energy scaling and computational cost and contrast it with CCSD(T). 
\begin{figure*}[t]
   \centering
   \includegraphics[width=0.95\textwidth]{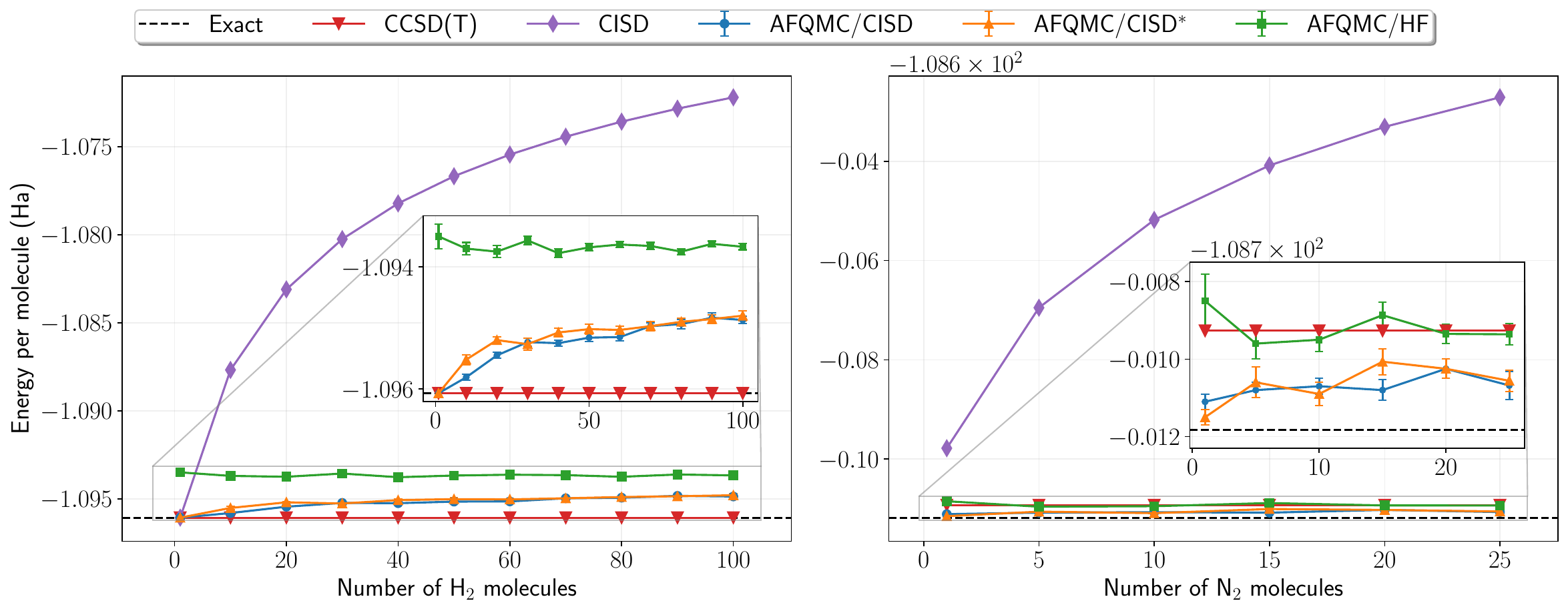}
   \caption{Energy per molecule for an increasing number of mutually non-interacting \ce{H2} (left) and \ce{N2} (right) molecules in the STO-6G basis. Insets show a zoomed-in view excluding the CISD energies. AFQMC/CISD\(^*\) refers to the trial state obtained variationally using CISD, as opposed to from CCSD amplitudes like in AFQMC/CISD.}
   \label{fig:extensivity}
\end{figure*}
\subsubsection{Size-extensivity}
Size-extensive methods ensure that the energy of a system composed of mutually non-interacting fragments (e.g., infinitely separated) equals the sum of the energies of the individual fragments. Therefore, for a system of widely separated molecules, the energy per molecule is independent of the number of molecules. Exact methods evidently have this property, as do some approximate methods like HF, CCSD, and CCSD(T). Ref. \citenum{lee2022twenty} demonstrated that AFQMC/HF energies also exhibit this behavior in the limit of a vanishing propagation time step. However, many correlated variational methods, CISD in particular, lack size-extensivity. Therefore, it is important to investigate the extensivity of AFQMC/CISD energies. Here, we provide a numerical analysis of this property. 

Fig. \ref{fig:extensivity} shows the energy per molecule for a system of well-separated \ce{H2} and \ce{N2} molecules using the STO-6G minimal basis set. We used bond distances close to equilibrium of \(d=2\) a.u. for \ce{H2} and and \(d=2.117\) a.u. for \ce{N2}. Since \ce{H2} is a two-electron system, CCSD and CCSD(T) yield exact energies for any number of non-interacting \ce{H2} molecules. We consider two CISD trials for AFQMC: one obtained from CCSD amplitudes, as we have done throughout this study, and the second obtained variationally using CISD. This second method is indicated as AFQMC/CISD\(^*\) in Fig. \ref{fig:extensivity}. 

Both AFQMC/CISD and AFQMC/CISD\(^*\) are only exact for a single \ce{H2} molecule since the trial CISD state is exact for this system. But as the number of molecules is increased, both flavors of AFQMC/CISD start deviating from the exact value, indicating a lack of size-extensivity. We find AFQMC/CISD energies to be slightly lower than AFQMC/CISD\(^*\) for a smaller number of molecules, but they seem to approach the same asymptote as the number of molecules is increased. The errors in both cases can be seen to be much smaller than that seen in CISD trial state energies. We also show AFQMC/HF energies in the figure, which can be seen to be size-extensive within the statistical error bars for the small time-step used here (\(\Delta t=0.005\) a.u.).  In the limit of an infinite number of molecules, the CISD energy reduces to the HF energy.\cite{szabo2012modern} Based on the presented numerical results, it is unclear if AFQMC/CISD and AFQMC/CISD\(^*\) energies converge to AFQMC/HF in this limit. It is worth noting that the CISD wave function does not collapse onto the HF state in this limit, with \(\inner{\psi_{\text{HF}}}{\psi_{\text{CISD}}} = 1/\sqrt{2}\), where both states are normalized.\cite{szabo2012modern} 

For well-separated \ce{N2} molecules, while CCSD(T) is no longer exact, it remains size-extensive. Again, although AFQMC/CISD and AFQMC/CISD\(^*\) are not exactly size-extensive, the deviations are small, and both energies have smaller errors than CCSD(T) for up to 25 molecules. Based on our calculations in this work, AFQMC/CISD is approximately variational and size-extensive, but not exactly so. We note that the lack of size-extensivity becomes less relevant if AFQMC/CISD is used within a local correlation approach.

\subsubsection{Computational scaling and cost}
\begin{figure}[t]
   \centering
   \includegraphics[width=0.45\textwidth]{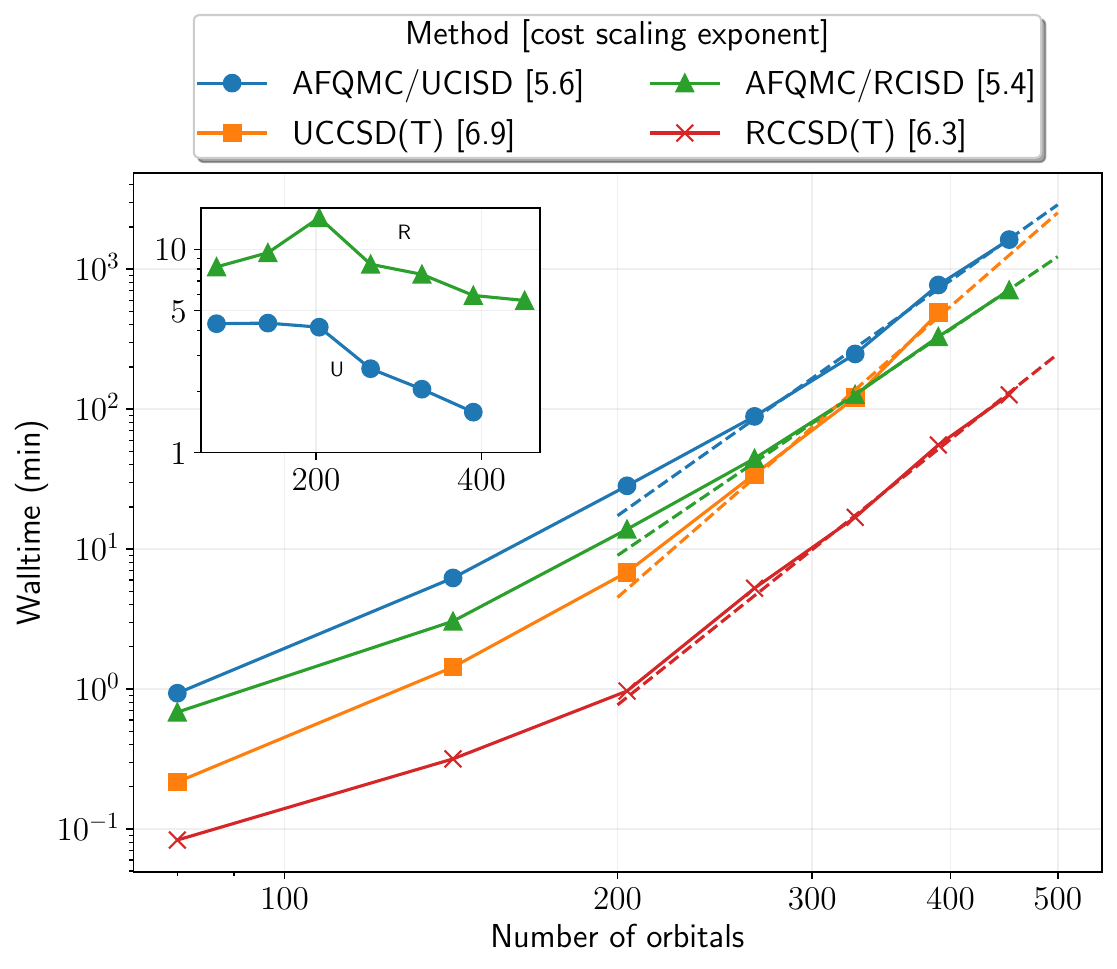}
   \caption{Energy calculation walltimes for trans-polyacetylene chains of increasing length (aug-cc-pVDZ basis) using AFQMC and CCSD(T). Both restricted and unrestricted versions of the two methods are shown. We performed AFQMC calculations using a GPU and CCSD(T) using a CPU (see text for details). We increased the number of AFQMC samples linearly with system size to keep the stochastic error fixed around 1 mH. The inset shows ratios of AFQMC and CCSD(T) walltimes.}
   \label{fig:times}
\end{figure}
Fig. \ref{fig:times} presents the walltimes (in minutes) for ground state energy calculations for trans-polyacetylene chains using the aug-cc-pVDZ basis set. These systems have a virtual-to-occupied orbital ratio of \(N_v / N_o \approx 7\). We performed both RHF and UHF-based calculations for each method on chains with up to sixteen carbon atoms. The AFQMC calculations were carried out on an NVIDIA A100 GPU with 40 GB of memory, while the CCSD(T) calculations were performed using PySCF on an AMD CPU node with 30 cores and 60 GB of memory. We note that our AFQMC code uses a negligible amount of disk space, whereas CCSD(T) requires a large amount of disk space for larger systems (up to 400 GB used in these calculations). Point group symmetry was not utilized in these calculations. Performing cost scaling analysis of a QMC method is complicated due to the stochastic error involved. For a fixed number of samples, the stochastic error roughly scales as \(\sqrt{N}\), where \(N\) represents the system size. To provide practical cost estimates, we increased the number of samples in AFQMC linearly with system size to maintain the stochastic error around 1 mH. A fifth of all samples at the beginning of the AFQMC run were discarded as equilibration samples. 

We determine empirical asymptotic cost scaling exponents by fitting walltime data against system size on a log-log scale for the three largest systems considered here. For RHF-based AFQMC/RCISD, the asymptotic scaling is \(O(N^{5.4})\), and for RCCSD(T) it is \(O(N^{6.3})\), roughly aligning with theoretical expectation. Performing CCSD(T) calculations for longer chains becomes challenging due to substantial disk space requirements. For smaller systems, AFQMC/RCISD calculations took approximately ten times longer than RCCSD(T). This ratio decreases asymptotically, reaching about 5 for the largest system. Similar asymptotic scaling is observed for UHF-based variants of these methods. The ratio of AFQMC/UCISD to UCCSD(T) walltimes decreases from around 5 for smaller systems to about 1.5 for the largest system. A common critique of QMC methods is their large prefactor in asymptotic scaling, making them computationally expensive despite favorable theoretical scaling. However, efficient GPU implementations significantly reduce this prefactor for AFQMC, making it competitive with CCSD(T) in practice. While GPU implementations of CC methods exist, they lack the embarrassing parallelizability of QMC methods. Further enhancements in the AFQMC implementation, such as mixed precision arithmetic, promise additional cost reductions.

\section{Conclusion}\label{sec:conclusion}
In this work, we explored the use of configuration interaction singles and doubles wave functions as trial states in AFQMC. Through benchmark calculations on various molecules, we established that AFQMC/CISD is generally more accurate than CCSD(T) for calculating ground state energies. Analyzing the cost scaling, we showed that AFQMC/CISD achieves this accuracy with a lower scaling of \(O(N^6)\) compared to the \(O(N^7)\) scaling of CCSD(T). We presented practical walltimes from our pilot GPU implementation, showing that AFQMC/CISD walltimes are comparable to a CPU-based CCSD(T) implementation. We found that this method is not strictly size extensive, but the deviations from exact extensivity were small for the cases we examined.

One possibility for future work is to lower the cost scaling of AFQMC/CISD by using low-rank and local correlation approaches. The use of tensor hypercontraction\cite{hohenstein2012tensor} for the CI coefficients promises a reduction in the cost scaling of energy calculations. Low-rank approximation techniques can also be employed for Hamiltonian integrals to reduce memory and walltime costs as demonstrated in previous AFQMC studies.\cite{malone2018overcoming,motta2019efficient,weber2022localized} Local correlation approaches, which offer nearly linear scaling with system size,\cite{riplinger2013efficient,rolik2011general} have recently been adapted to AFQMC using local natural orbitals.\cite{kurian2023toward} Using this technique for AFQMC/CISD would provide a significant reduction in the computational cost. 

Other potential research directions include the calculation of other properties and nuclear gradients within a response formalism. We have already shown that the differentiable framework developed in Ref. \citenum{mahajan2023response} produces accurate estimates of dipole moments and other one-body properties. Its extension to nuclear gradients, combined with the accuracy of AFQMC/CISD across potential energy surfaces, could become a valuable tool for molecular dynamics simulations. Another promising avenue is the calculation of excited states within AFQMC/CISD. We plan to pursue these directions in future work.

\section*{Data availability}
The code used for AD-AFQMC calculations is available in a public GitHub repository at Ref. \citenum{dqmc_code}. Raw energies and input scripts are also available in a public repository at Ref. \citenum{afqmc_files}.

 \section*{Acknowledgements}
 S.S. and J.S.K were partially supported by NSF CHE-2145209. A.M. and D.R.R. were partially supported by NSF CHE-2245592. J.H.T. acknowledges support from the SMU Moody School of Graduate and Advanced Studies. D.A.M. was partially supported by Department of Energy (Grant No. DE-SC0022893). This work used the Delta system at the National Center for Supercomputing Applications through allocation CHE230028 from the Advanced Cyberinfrastructure Coordination Ecosystem: Services and Support (ACCESS) program, which is supported by National Science Foundation grants \#2138259, \#2138286, \#2138307, \#2137603, and \#2138296. Some calculations were performed on computing resources provided by SMU’s O’Donnell Data Science and Research Computing Institute.

\appendix
\section{Computational details for CISD trial states}\label{app:cisd}
In this section, we derive the force bias and local energy expressions for the CISD trial based on an RHF reference state. Analogous expressions can be derived for trial states based on a UHF reference. The CISD trial is given by
\begin{equation}
   \ket{\psi_T} = \left(1 + c^{t}_{p}a_{t\sigma}^{\dagger}a_{p\sigma} + \frac{1}{2}c^{tu}_{pq}a_{t\sigma}^{\dagger}a_{u\lambda}^{\dagger}a_{q\lambda}a_{p\sigma}\right) \ket{\phi_0},  
\end{equation}
where we have used the convention of summing over repeated indices here and in the following. The overlap of the CISD trial with a walker determinant \(\ket{\phi}\) is evaluated using the ratio
\begin{equation}
   \frac{\inner{\psi_T}{\phi}}{\inner{\phi_0}{\phi}} = 1 + 2c^{t}_{p}G_{t}^{p} + c^{tu}_{pq}(2G_{t}^{p}G_{u}^{q} - G_{u}^{p}G_{t}^{q}),\label{eq:overlap_ratio}
\end{equation}
where \(\inner{\phi_0}{\phi}\) is given by a determinant. It is convenient to separate the force bias into zero, one, and two body contributions as
\begin{equation}
   v_{\gamma} = L^{\gamma}_{ij}\frac{\expecth{\psi_T}{a_{i\sigma}^{\dagger}a_{j\sigma}}{\phi}}{\inner{\phi_0}{\phi}}\frac{\inner{\phi_0}{\phi}}{\inner{\psi_T}{\phi}} = \left(v_{\gamma}^0 + v_{\gamma}^1 + v_{\gamma}^2\right)\frac{\inner{\phi_0}{\phi}}{\inner{\psi_T}{\phi}}.
\end{equation}
These contributions are evaluated as follows:
\begin{equation}
   v_{\gamma}^0 = L^{\gamma}_{ij}\frac{\expecth{\phi_0}{a_{i\sigma}^{\dagger}a_{j\sigma}}{\phi}}{\inner{\phi_0}{\phi}} = 2L^{\gamma}_{ij}G_{j}^{i},
\end{equation}
\begin{equation}
   \begin{split}
      v_{\gamma}^1 &= L^{\gamma}_{ij}c^{t}_{p}\frac{\expecth{\phi_0}{a_{p\sigma}^{\dagger}a_{t\sigma}a_{i\lambda}^{\dagger}a_{j\lambda}}{\phi}}{\inner{\phi_0}{\phi}}\\
      &= L^{\gamma}_{ij}c^{t}_{p} \begin{vmatrix}
      G_{j}^{i}\delta_{\lambda\lambda} & \mathcal{G}_{t}^{i}\delta_{\sigma\lambda} \\[0.5em]
      G_{j}^{p}\delta_{\lambda\sigma} & G_{j}^{i}\delta_{\sigma\sigma}   
      \end{vmatrix}\\
      &= 2L^{\gamma}_{ij}c^{t}_{p}(2G_{t}^{p}G_{j}^{i} - G_{j}^{p}\mathcal{G}_{t}^{i}),
   \end{split}
\end{equation}
and
\begin{equation}
   \begin{split}
      v_{\gamma}^2 &= \frac{1}{2}L^{\gamma}_{ij}c^{tu}_{pq}\frac{\expecth{\phi_0}{a_{p\sigma}^{\dagger}a_{q\lambda}^{\dagger}a_{u\lambda}a_{t\sigma}a_{i\kappa}^{\dagger}a_{j\kappa}}{\phi}}{\inner{\phi_0}{\phi}}\\
      &= \frac{1}{2}L^{\gamma}_{ij}c^{tu}_{pq} \begin{vmatrix}
      G_{j}^{i}\delta_{\kappa\kappa} & \mathcal{G}_{t}^{i}\delta_{\kappa\sigma} & \mathcal{G}_{u}^{i}\delta_{\kappa\lambda} \\[0.5em]
      G_{j}^{p}\delta_{\sigma\kappa} & G_{t}^{p}\delta_{\sigma\sigma} & G_{u}^{p}\delta_{\sigma\lambda} \\[0.5em]
      G_{j}^{q}\delta_{\lambda\kappa} & G_{t}^{q}\delta_{\lambda\sigma} & G_{u}^{q}\delta_{\lambda\lambda}
      \end{vmatrix}\\
      &= 2L^{\gamma}_{ij}c^{tu}_{pq}\left[ G_{j}^{i}\left(2G_{t}^{p}G_{u}^{q}-G_{u}^{p}G_{t}^{q}\right)\right.\\
      &\qquad\qquad\qquad\qquad \left.-\mathcal{G}_{t}^{i}\left(2G_{j}^{p}G_{u}^{q} - G_{u}^{p}G_{j}^{q}\right)\right].
   \end{split}
\end{equation}
The local energy is given by
\begin{equation}
   E_L = \frac{\expecth{\psi_T}{H}{\phi}}{\inner{\phi_0}{\phi}}\frac{\inner{\phi_0}{\phi}}{\inner{\psi_T}{\phi}} = \left({}^0E_L + {}^1E_L + {}^2E_L\right)\frac{\inner{\phi_0}{\phi}}{\inner{\psi_T}{\phi}},
\end{equation}
where \({}^0E_L\), \({}^1E_L\), and \({}^2E_L\) are explicitly described below. The zero body term \({}^0E_L\) is the same as the overlap ratio in Eq. \ref{eq:overlap_ratio}. The one-body term \({}^1E_L\) can be evaluated similarly to force bias. The two-body term is again more convenient to evaluate by separating it into zero, one, and two-body contributions:
\begin{equation}
   {}^2E_L = \frac{1}{2}L^{\gamma}_{ij}L^{\gamma}_{kl}\frac{\expecth{\psi_T}{a_{i\sigma}^{\dagger}a_{k\lambda}^{\dagger}a_{l\lambda}a_{j\sigma}}{\phi}}{\inner{\phi_0}{\phi}} = {}^2E_L^0 + {}^2E_L^1 + {}^2E_L^2.
\end{equation}
These contributions are expressed as follows:
 \begin{align}
   \begin{split}
      {}^2E_L^0 &= \frac{1}{2}L^{\gamma}_{ij}L^{\gamma}_{kl}\frac{\expecth{\phi_0}{a_{i\sigma}^{\dagger}a_{k\lambda}^{\dagger}a_{l\lambda}a_{j\sigma}}{\phi}}{\inner{\phi_0}{\phi}}\\
      &= \frac{1}{2}L^{\gamma}_{ij}L^{\gamma}_{kl}\begin{vmatrix}
      G_{j}^{i}\delta_{\sigma\sigma} & G_{l}^{i}\delta_{\sigma\lambda} \\[0.5em]
      G_{j}^{k}\delta_{\lambda\sigma} & G_{l}^{k}\delta_{\lambda\lambda}
      \end{vmatrix}\\
      &= L^{\gamma}_{ij}L^{\gamma}_{kl}(2G_{j}^{i}G_{l}^{k} - G_{l}^{i}G_{j}^{k}),
   \end{split}
\end{align}
\begin{align}
   \begin{split}
      {}^2E_L^1 &= \frac{1}{2}L^{\gamma}_{ij}L^{\gamma}_{kl}c^{t}_{p}\frac{\expecth{\phi_0}{a_{p\sigma}^{\dagger}a_{t\sigma}a_{i\lambda}^{\dagger}a_{k\mu}^{\dagger}a_{l\mu}a_{j\lambda}}{\phi}}{\inner{\phi_0}{\phi}}\\
      &= \frac{1}{2}L^{\gamma}_{ij}L^{\gamma}_{kl}c^{t}_{p} \begin{vmatrix}
      G_{j}^{i}\delta_{\lambda\lambda} & G_{l}^{i}\delta_{\lambda\mu} & \mathcal{G}_{t}^{i}\delta_{\lambda\sigma} \\[0.5em]
      G_{j}^{k}\delta_{\mu\lambda} & G_{l}^{k}\delta_{\mu\mu} & \mathcal{G}_{t}^{k}\delta_{\mu\sigma} \\[0.5em]
      G_{j}^{p}\delta_{\sigma\lambda} & G_{l}^{p}\delta_{\sigma\mu} & G_{t}^{p}\delta_{\sigma\sigma}
      \end{vmatrix}\\
      &= 2L^{\gamma}_{ij}L^{\gamma}_{kl}c^{t}_{p}\left[G_{t}^{p}\left(2G_{j}^{i}G_{l}^{k} - G_{l}^{i}G_{j}^{k}\right)\right.\\
      &\qquad\qquad\qquad\qquad \left.-G_{l}^{p}(2G_{j}^{i}\mathcal{G}_{t}^{k} - G_{j}^{k}\mathcal{G}_{t}^{i})\right],
   \end{split}
\end{align}
and
\begin{widetext}
 \begin{align}
   \begin{split}
       {}^2E_L^2 &= \frac{1}{4}L^{\gamma}_{ij}L^{\gamma}_{kl}c^{tu}_{pq}\frac{\expecth{\phi_0}{a_{p\sigma}^{\dagger}a_{q\nu}^{\dagger}a_{u\nu}a_{t\sigma}a_{i\lambda}^{\dagger}a_{k\mu}^{\dagger}a_{l\mu}a_{j\lambda}}{\phi}}{\inner{\phi_0}{\phi}}\\
      &= \frac{1}{4}L^{\gamma}_{ij}L^{\gamma}_{kl}c^{tu}_{pq} \begin{vmatrix}
      G_{j}^{i}\delta_{\lambda\lambda} & G_{l}^{i}\delta_{\lambda\mu} & \mathcal{G}_{t}^{i}\delta_{\lambda\sigma} & \mathcal{G}_{u}^{i}\delta_{\lambda\nu} \\[0.5em]
      G_{j}^{k}\delta_{\mu\lambda} & G_{l}^{k}\delta_{\mu\mu} & \mathcal{G}_{t}^{k}\delta_{\mu\sigma} & \mathcal{G}_{u}^{k}\delta_{\mu\nu} \\[0.5em]
      G_{j}^{p}\delta_{\sigma\lambda} & G_{l}^{p}\delta_{\sigma\mu} & G_{t}^{p}\delta_{\sigma\sigma} & G_{u}^{p}\delta_{\sigma\nu} \\[0.5em]
      G_{j}^{q}\delta_{\nu\lambda} & G_{l}^{q}\delta_{\nu\mu} & G_{t}^{q}\delta_{\nu\sigma} & G_{u}^{q}\delta_{\nu\nu}
      \end{vmatrix}\\
      &= L^{\gamma}_{ij}L^{\gamma}_{kl}c^{tu}_{pq}\left[(2G_{j}^{i}G_{l}^{k} - G_{l}^{i}G_{j}^{k})(2G_{t}^{p}G_{u}^{q}-G_{t}^{q}G_{u}^{p})-(2G_{l}^{k}\mathcal{G}_{t}^{i} - G_{l}^{i}\mathcal{G}_{t}^{k})(2G_{j}^{p}G_{u}^{q} - G_{u}^{p}G_{j}^{q})\right.\\
      &\hspace*{30em}\left.+(G_{j}^{p}G_{l}^{q} - G_{j}^{q}G_{l}^{p})(\mathcal{G}_{t}^{i}\mathcal{G}_{u}^{k}-\mathcal{G}_{t}^{k}\mathcal{G}_{u}^{i})\right].
   \end{split}
\end{align}
\end{widetext}
\begin{figure*}[t]
   \centering
   \includegraphics[width=0.8\textwidth]{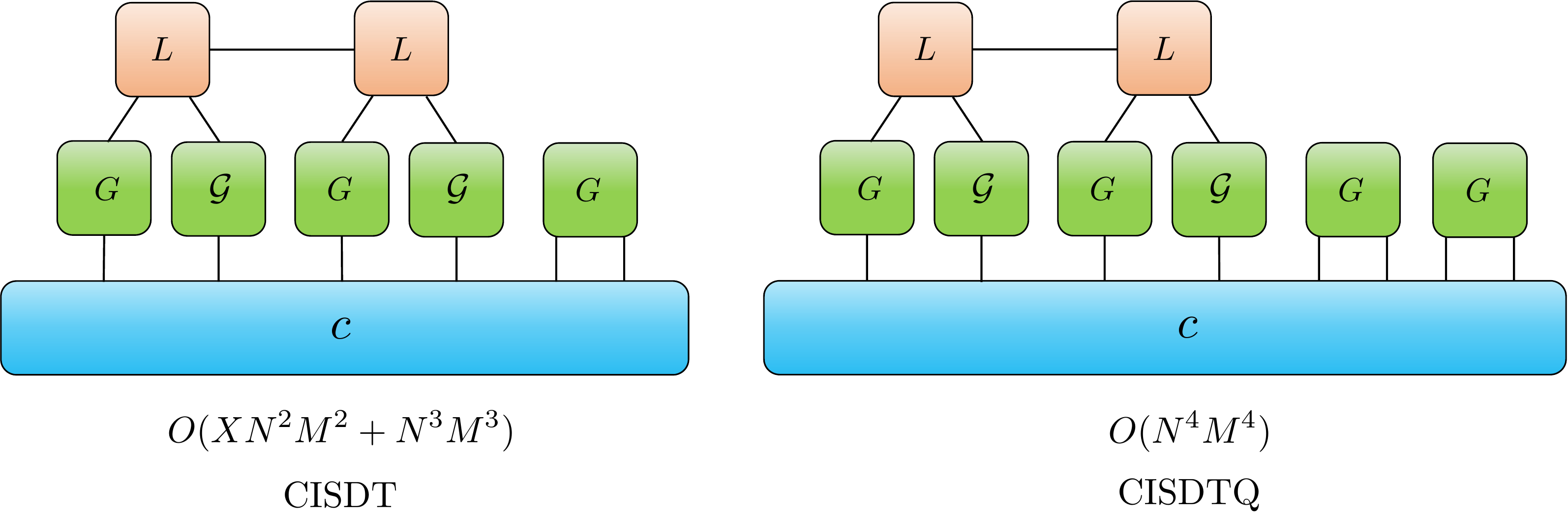}
   \caption{Tensor diagrams for the most expensive terms in the evaluation of local energy for CISDT and CISDTQ trial states. Computational cost scaling is shown below each diagram.}
   \label{fig:cisdtq}
\end{figure*}
Expressions for higher-order trial states can be similarly derived. Tensor diagrams for the most expensive terms in the calculation of local energy for CISDT and CISDTQ trial states are shown in Fig. \ref{fig:cisdtq}.

\section{Atomic energies}\label{app:atomic}
\begin{figure}
   \centering
   \includegraphics[width=0.45\textwidth]{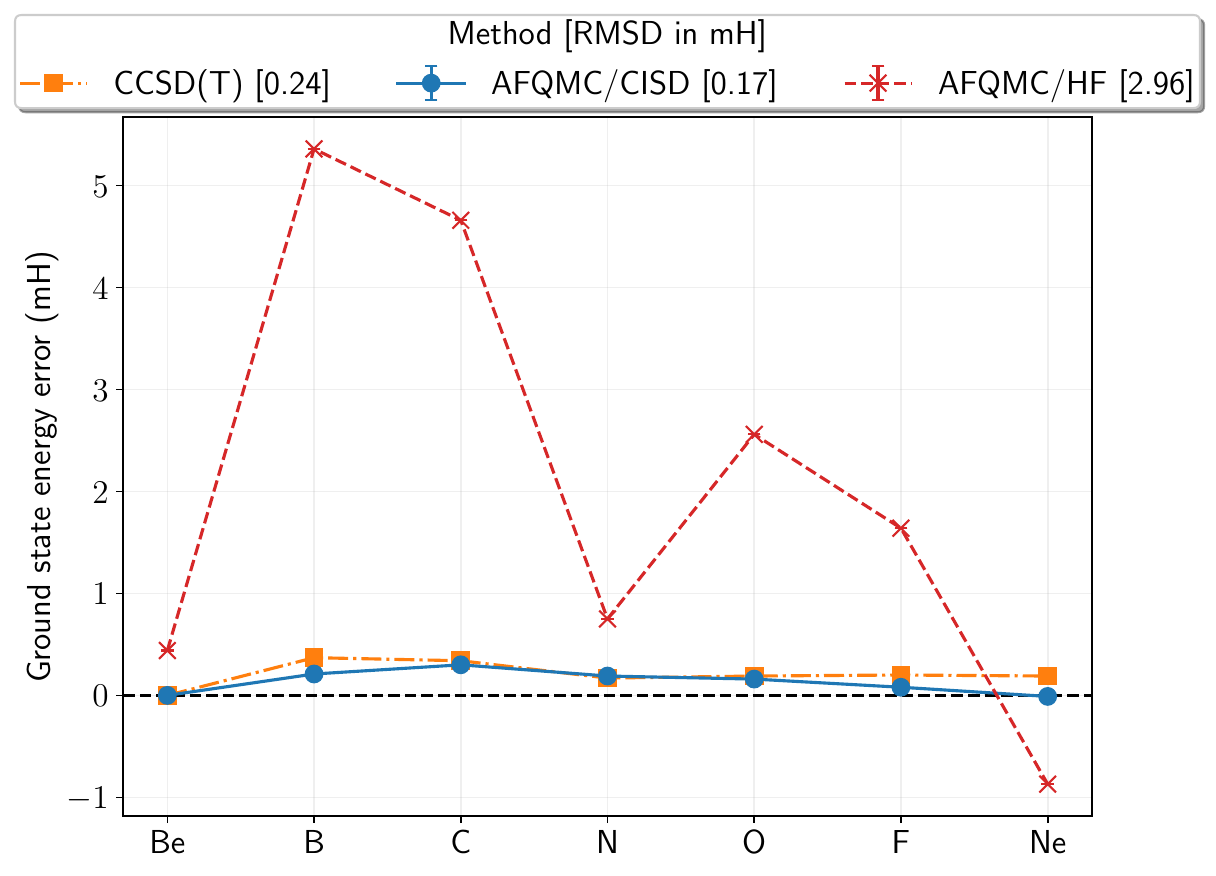}
   \caption{Errors in the ground state energies for the first row atoms using the cc-pVDZ basis set.}
   \label{fig:atomic}
\end{figure}

Fig. \ref{fig:atomic} shows the errors in the ground state atomic energies of first-row elements. We used the cc-pVDZ basis set and exact full CI energies as the reference. AFQMC/HF shows unexpectedly large errors for this seemingly simple setting of ground state atomic electronic structure, aligning with observations in Ref. \citenum{lee2022twenty}. The addition of single and double excitations in the trial in AFQMC/CISD corrects this behavior almost entirely, resulting in energies slightly better than CCSD(T). This highlights the critical role of these excitations in accurately capturing the ground state sign structure.

%
   
\end{document}